\documentclass[preprint]{aastex}

\begin{document}

\shorttitle{Submillimeter Constraints on Disk Structure}

\shortauthors{Andrews \& Williams}

\title{High Resolution Submillimeter Constraints on Circumstellar Disk Structure}

\author{Sean M. Andrews and Jonathan P. Williams}

\affil{Institute for Astronomy, University of Hawaii, 2680 Woodlawn Drive, Honolulu, HI 96822}
\email{andrews@ifa.hawaii.edu, jpw@ifa.hawaii.edu}

\begin{abstract}
We present a high spatial resolution submillimeter continuum survey of 24 
circumstellar disks in the Taurus-Auriga and Ophiuchus-Scorpius star formation 
regions using the SMA.  In the context of a simple model, we use broadband 
spectral energy distributions and submillimeter visibilities to derive 
constraints on some basic parameters that describe the structure of these 
disks.  For the typical disk in the sample we infer a radial surface density 
distribution $\Sigma_r \propto r^{-p}$ with a median $p \approx 0.5$, although 
consideration of the systematic effects of some of our assumptions suggest that 
steeper distributions with $p \approx 0.7-1.0$ are more reasonable.  The 
distribution of the outer radii of these disks shows a distinct peak at $R_d 
\approx 200$\,AU, with only a few cases where the disk emission is completely 
unresolved.  Based on these disk structure measurements, the mass accretion 
rates, and the typical spectral and spatial distributions of submillimeter 
emission, we show that the observations are in good agreement with similarity 
solutions for steady accretion disks that have a viscosity parameter $\alpha 
\approx 0.01$.  We provide new estimates of the spectral dependence of the disk 
opacity $\kappa_{\nu} \propto \nu^{\beta}$ with a median $\beta \approx 0.7$, 
corrected for optically thick emission.  This typical value of $\beta$ is 
consistent with model predictions for the collisional growth of solids to 
millimeter size scales in the outer disk.  Although direct constraints on 
planet formation in these disks are not currently available, the extrapolated 
density distributions inferred here are substantially shallower than those 
calculated based on the solar system or extrasolar planets and typically used 
in planet formation models.  It is possible that we are substantially 
underestimating disk densities due to an incomplete submillimeter opacity 
prescription.
\end{abstract}

\section{Introduction}
Circumstellar disks play integral roles in early stellar evolution and the 
genesis of planetary systems.  As the gas and dust reservoirs that contain the 
raw material for building planets, these disks provide a snapshot of the planet 
formation process.  Any model of this process is necessarily dependent on the 
distribution and composition of the progenitor disk material 
\citep[e.g.,][]{pollack96,inaba03,boss05,durisen05}.  In principle, 
observations related to the structure and content of these disks can be used to 
constrain the timescales and mechanisms involved in building planets.  Key 
insights into the origins of planetary systems can also be determined based on 
the dynamical and physical properties of extrasolar planets 
\citep[e.g.,][]{marcy00,udry06} and their host stars 
\citep{santos01,fischer05}, as well as via the internal structure and 
composition of the giant planets \citep{lunine04,guillot05} and various aspects 
of the populations of smaller bodies \citep[e.g.,][]{luu02} in the solar 
system.  Studies of the ancestral circumstellar disks and their descendent 
planetary systems approach the topic of planet formation from opposite 
directions in time.  

Observations that are sensitive to the structure of disks are also useful for 
constraining the internal mechanisms that govern their evolution.  For example, 
the spatial density distribution and size of a disk can be combined with the 
mass accretion rate onto the central star 
\citep[e.g.,][]{valenti93,hartigan95,gullbring97,muzerolle03a} to estimate the 
disk viscosity \citep[e.g.,][]{hartmann98}.  This viscosity is thought to be 
generated by the magnetorotational instability \citep[][but see Hartmann et 
al.~2006]{balbus91} and, along with gravity and the conservation of angular 
momentum, dictates the structural evolution of disk material 
\citep{lyndenbell74,lin82,hartmann98,hueso05}.  As a second example, the 
spectral dependence of the disk opacity can be used to estimate the size 
distribution of solid particles in the disk, thereby tracing the growth of dust 
grains to the earliest planetesimals 
\citep{miyake93,dalessio01,dalessio06,draine06}.  This collisional 
agglomeration of disk material plays a critical role in both the evolution of 
the disk and the planet formation process.

While there is clearly significant motivation to study the physical structure 
of circumstellar disks, interpreting the observational data in this context is 
a challenge.  The focus of this paper is on observations of continuum emission 
from these disks, particularly at submillimeter wavelengths.\footnote{For 
convenience, we define ``submillimeter" to broadly incorporate wavelengths 
between a few hundred microns and a few millimeters.}  This thermal emission is 
reprocessed starlight from irradiated dust grains.  The spectral energy 
distribution (SED) of this emission is determined by a range of disk radii 
which have different temperature and density conditions.   As a consequence, 
submillimeter continuum observations play an important role in constraining 
disk structure properties for three primary reasons.  First, the great majority 
of the mass and volume of a typical disk (with radius $\gtrsim$ 100\,AU) will 
be relatively cold, therefore emitting the bulk of the continuum at these 
wavelengths.  Second, there should be substantial submillimeter emission in the 
extended outer regions of a disk, making spatially resolved observations with 
an interferometer possible.  And third, much of the submillimeter emission is 
thought to be optically thin \citep[e.g.,][]{bscg90}, meaning measurements of 
its spatial and spectral distributions can be used to readily infer the mass 
distribution and opacity in the disk.

The pioneering high resolution observations of submillimeter continuum emission 
from circumstellar disks were directed at measuring their sizes and 
orientations \citep[e.g.,][]{keene90,lay94}.  As the technology and data 
quality improved, focus shifted to exploring the radial structure of a few 
resolved disks \citep[e.g., HL Tau;][]{mundy96,wilner96}, and then in 
particular to determining constraints on their sizes and density distributions 
\citep[e.g.,][]{dutrey96,lay97,akeson98,akeson02,wilner00,kitamura02}.  A set 
of complementary studies used spatially resolved line emission from molecular 
gas phase tracers like CO to work toward the same end 
\citep[e.g.,][]{dutrey98,guilloteau98,guilloteau99}.  More recently, a few of 
these disks have been examined under the auspices of more sophisticated models 
\citep{dutrey98,dalessio01,calvet02,dartois03} to place ever more detailed 
constraints on properties like structure and chemistry 
\citep{wilner03,qi03,qi04,qi06,dutrey06}, as well as the signatures of grain 
growth \citep[e.g.,][]{wilner05}.

In this paper, we utilize a survey of high spatial resolution submillimeter 
continuum observations and a simple model to place constraints on the physical 
structure of circumstellar disks.  Measurements of molecular line emission 
(rotational transitions of CO) from these same data will be presented in a 
separate paper.  The observations, data reduction, and basic sample properties 
are introduced in \S 2.  The modeling procedure is described in detail in \S 3, 
and basic constraints on disk structure parameters are given in \S 4.  A 
discussion in \S 5 aims to synthesize the results in the contexts of disk 
evolution and planet formation, with some comments on the prospects for future 
work on this topic.  The Appendix contains additional comments on individual 
disks in the sample.

\section{Observations and Data Reduction}
Interferometric observations of 24 young star/disk systems were conducted with 
the Submillimeter Array \citep[SMA;][]{ho04} on the summit of Mauna Kea, 
Hawaii.  The SMA consists of eight 6\,m antennas which can be placed on 24 pads 
across a relatively flat valley at an altitude of $\sim$4070\,m.  All targets 
were observed in one of two different compact array configurations with 
baselines up to $\sim$150\,m (the ``C1" configuration, used before 2005 May) or 
$\sim$70\,m (the ``C2" configuration, used after 2005 May).  Many of the 
targets were also observed in an extended (E) configuration of the array with 
maximum baseline lengths of $\sim$200\,m.  

Double sideband receivers were tuned to an intermediate frequency (IF) of 
either 225.494, 340.758, or 349.930\,GHz (1330, 880, or 857\,$\mu$m, 
respectively).  Each sideband provides 2\,GHz of bandwidth, centered $\pm 
5$\,GHz from the IF.  The standard correlator setup adopted in this survey 
provides 24 partially overlapping basebands of 104\,MHz width in each sideband, 
with a 0.8125\,MHz channel spacing in each baseband.  The highest frequency 
observations ($\nu_{\rm{IF}} = 349.930$\,GHz) used a slightly different 
correlator setup to accomodate higher spectral resolution in some basebands, 
leading to a slightly reduced total continuum bandwidth.

Most of the observations interleaved two targets and two quasars (complex gain 
calibrators) in an alternating pattern, with 20 minutes on a target and then 10 
minutes on a quasar.  Additional calibrators were observed at the beginning and 
end of a night.  The quasars used for complex gain calibration were 3C\,111 and 
3C\,84 for Taurus-Auriga targets,\footnote{On 2005 September 9, J0528+134 
replaced 3C\,84 as a calibrator.} and J1733$-$130 and J1743$-$038 or 
J1517$-$243 and J1626$-$298 for Ophiuchus-Scorpius targets.  Planets (Uranus, 
Jupiter, Saturn), satellites (Titan, Callisto), and quasars (3C\,454.3, 
3C\,273, 3C\,279) were observed as passband and absolute flux calibrators 
depending on their availability and the array configuration.  Observing 
conditions were generally excellent.  Data were obtained with $\lesssim$1.2 and 
3\,mm of precipitable water vapor for high and low frequencies, respectively, 
with system temperatures in the range of 100$-$400\,K.  A summary of basic 
observational information is given in Table \ref{obssum_table}.

The data were edited and calibrated using MIR,\footnote{\url{http://cfa-www.harvard.edu/$\sim$cqi/mircook.html}} an IDL-based software package originally 
developed for the OVRO array and adapted by the SMA group.  After appropriate 
editing, calibration of the passband response for each baseband was determined 
using a bright planet or quasar.  Broadband continuum channels in each sideband 
were generated by averaging the central 82\,MHz in all line-free basebands.  
The baseline-based complex gain response of the system was calibrated using one 
or both of the quasars interleaved with the targets.  Absolute flux calibration 
was performed based on either planets/satellites (Uranus, Titan, or Callisto) 
and/or routinely-monitored quasars (e.g., 3C\,454.3).  Typical systematic 
uncertainties in the absolute flux scale of $\sim$10$-$15\% were determined 
based both on the uncertainties of the planetary emission models or quasar flux 
densities and the level of agreement between various methods of performing the 
calibrations (data from 2004 had absolute flux calibration uncertainties at the 
$\sim$20\% level).  

The standard tasks of Fourier inversion, deconvolution with the CLEAN 
algorithm, and restoration with the synthesized beam were conducted with the 
MIRIAD software package.  All continuum maps were created with natural 
weighting in the Fourier plane.  Synthesized beam parameters are given in Table 
\ref{obssum_table}.  Maps of the gain calibrators were checked against one 
another to determine the effects of pointing errors, seeing, and any small 
baseline errors.  In general, these effects lead to positional uncertainties on 
the order of 0\farcs1 or less, with observations in 2004 slightly worse.  The 
J2000 phase center coordinates were chosen to coincide with the stellar 
positions, determined from the \emph{Two Micron All-Sky Survey} (2MASS) Point 
Source Catalog astrometry \citep{cutri03}.

\begin{figure}
\epsscale{1.0}
\plotone{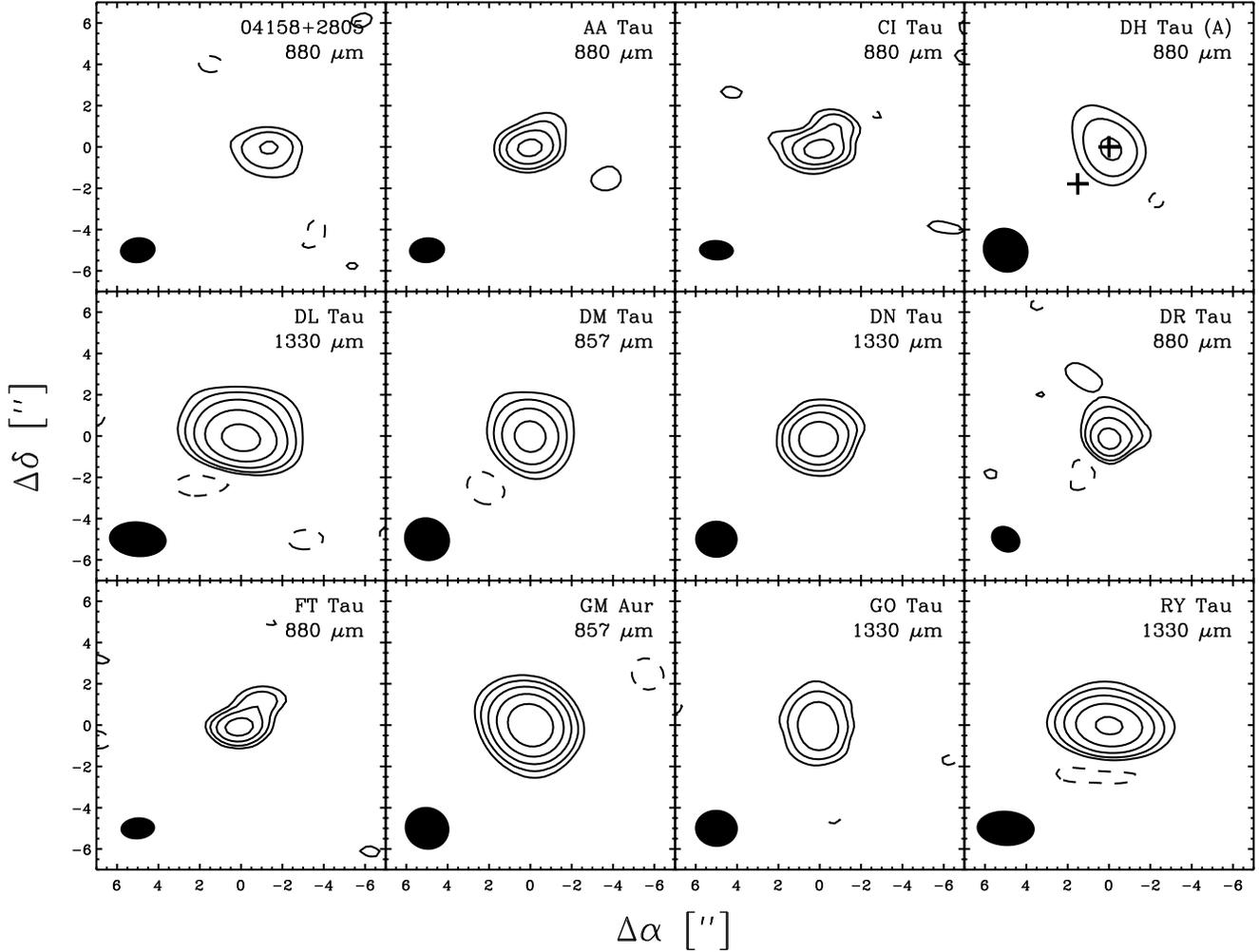}
\vspace{0.5cm}
\figcaption{Aperture synthesis images of submillimeter continuum emission for 
the sample disks in Taurus-Auriga.  The axes are offsets in arcseconds.  The 
FWHM dimensions and orientations of the naturally-weighted synthesized beams 
are shown in the lower left corner of each panel.  Contours begin at 
4\,$\sigma$ and step in factors of two in intensity.  Multiple star systems 
have individual components marked with crosses.  The component that exhibits 
the continuum emission is denoted by the object label at the top right. 
\label{contmaps_Tau}}
\end{figure}

\begin{figure}
\epsscale{1.0}
\plotone{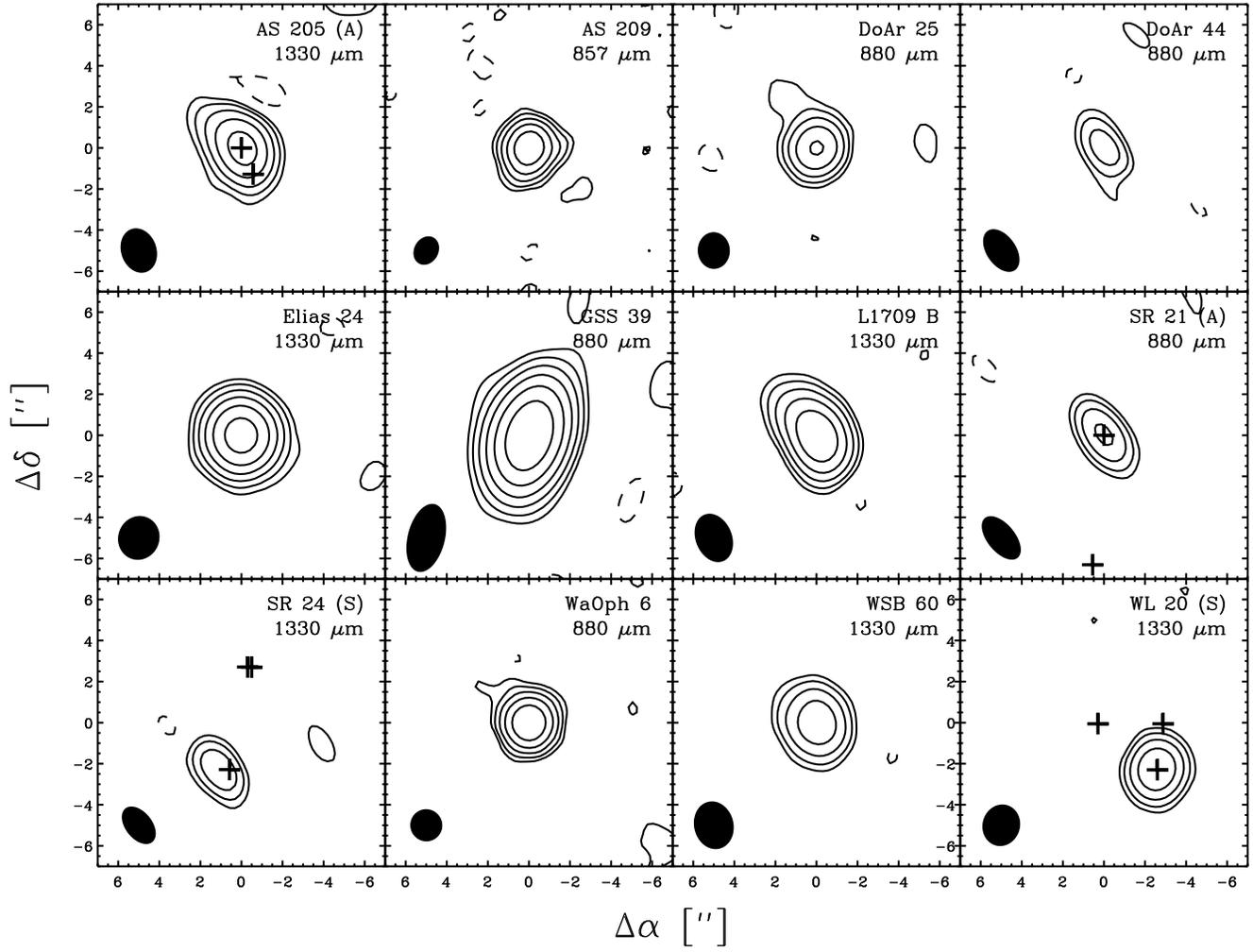}
\vspace{0.5cm}
\figcaption{Same as Fig.~\ref{contmaps_Tau} for the sample disks in 
Ophiuchus-Scorpius.  \label{contmaps_Oph}}
\end{figure}

Continuum emission maps of the sample objects are shown in Figures 
\ref{contmaps_Tau} and \ref{contmaps_Oph}.  The axes mark offsets in 
arcseconds, and the synthesized beam size and shape are shown in the lower left 
corner of each panel.  Continuum flux densities for each source were determined 
by summing the emission within the 2\,$\sigma$ contour, with the rms noise 
determined in an emission-free box within $\pm$20\arcsec\ from the phase 
center.  The continuum flux densities and statistical errors are listed in 
Table \ref{cont_table}, along with FWHM source dimensions and orientations 
determined from elliptical Gaussian fits to the visibilities.  Maps of multiple 
star systems have individual stellar positions marked with crosses, with the 
component exhibiting continuum emission clearly labeled in Table 
\ref{cont_table} and Figures \ref{contmaps_Tau}$-$\ref{contmaps_Oph}.  In most 
cases, the interferometer has recovered all of the continuum flux observed with 
single-dish data \citep{bscg90,andre94,nurnberger98,andrews05}.  Several of the 
sources, particularly those in the Ophiuchus clouds, have SMA flux densities 
lower than those determined with a single-dish telescope.  Presumably the 
interferometric fluxes are lower due to the spatial filtering of extended 
cloud/envelope emission in these cases (see the Appendix).

The targets for this study were selected primarily by their single-dish 
submillimeter flux densities to ensure fairly high signal-to-noise ratios for 
the developing SMA.  This criterion introduces a significant bias to the 
sample, as the brightest submillimeter disks are not necessarily 
representative.  The median 850\,$\mu$m flux density for a disk in the sample 
of \citet{andrews05} is several times smaller than the SMA sample median.  The 
sample is split evenly among disks in the Taurus-Auriga complex \citep[$d 
\approx 140$\,pc;][]{elias78} and the Ophiuchus-Scorpius region \citep[$d 
\approx 160$\,pc;][]{degeus89}.  Despite the limitations of the primary 
selection criterion, we attempted to compose a sample with a wide range of 
stellar and disk properties.  The stars in the sample basically span the T 
Tauri spectral type range, from early G to middle M types (corresponding to 
$M_* \approx 0.1-2$\,M$_{\odot}$), while the disks exhibit a broad scope of the 
standard characteristics attributed to accretion and/or excess photospheric 
emission.  A significant fraction of the sample targets ($\sim$20\%; 5/24) are 
known multiple star systems.  Some basic properties of the targets are compiled 
in Table \ref{stars_table}.  

\section{Models and Disk Structure Constraints}
The spatial distribution and SED of thermal continuum emission are the primary 
data available for determining the structural properties of circumstellar dust 
disks.  Reasonable estimates of the radial temperature distribution and total 
disk mass can be made from the mid-infrared SED and submillimeter photometry, 
respectively \citep[e.g.,][]{bscg90,andrews05}.  However, it is not possible to 
place constraints on either the density distribution or size of a disk without 
spatially resolved data.  In principle, the combination of a complete SED and a 
spatially resolved image can be used to determine these parameters with a 
direct fit to a disk model.

One simple case is the ``flat" disk model, which allows for a continuous radial 
distribution of material with power-law forms for the temperature and surface 
density \citep{adams87}.  This model is able to reproduce the SEDs and 
submillimeter images of typical circumstellar disks quite well, and serves as a 
reasonable approximation for the more detailed structure of a realistic 
accretion disk.  More sophisticated models have been advanced to address the 
observational and theoretical shortcomings of this model \citep[for reviews, 
see][]{dullemond06,dutrey06}.  For example, a disk subject to hydrostatic 
equilibrium in the vertical direction is flared, and therefore has an increased 
illuminated surface area at radii corresponding to temperatures that produce 
mid- and far-infrared emission \citep{kenyon87,chiang97,chiang99,dullemond02}.  
However, because most of the submillimeter emission is generated near the 
midplane of the disk, this vertical flaring does not have a strong effect on 
the emission at those wavelengths \citep[e.g.,][]{chiang97}.  While these more 
complex models illuminate key properties of disks, the computing power required 
to estimate their parameters with a minimization technique can be formidable.  
Based on this fact and the size and quality of the sample presented here, we 
focus on interpreting these data in terms of the comparatively simple flat disk 
description.  

\subsection{The Flat Disk Model}
In the flat disk model, photospheric excess emission is generated by thermal 
reprocessing of starlight by a geometrically thin dust disk with radial 
temperature ($T_r$) and surface density ($\Sigma_r$) profiles described by 
power-laws:
\begin{equation}
T_r = T_1 \left(\frac{r}{1\,\rm{AU}}\right)^{-q}
\end{equation}
\begin{equation}
\Sigma_r = \Sigma_5 \left(\frac{r}{5\,\rm{AU}}\right)^{-p}
\end{equation}
where $T_1$ is the temperature at 1\,AU and $\Sigma_5$ is the surface density 
at 5\,AU (the normalization of $\Sigma_r$ is selected for comparison to 
Jupiter).  By further assuming an opacity spectrum that is a power-law in 
frequency and independent of radius, 
\begin{equation}
\kappa_{\nu} = \kappa_0 \left(\frac{\nu}{\nu_0}\right)^{\beta},
\end{equation}
the flux density ($F_{\nu}$) at any frequency can be determined by summing the 
thermal emission from annuli weighted by the optical depth ($\tau_{\nu,r} = 
\kappa_{\nu} \Sigma_r$), 
\begin{equation}
F_{\nu} = \frac{\cos{i}}{d^2} \int^{R_d}_{r_0} B_{\nu}(T_r) (1-e^{-\tau_{\nu, r}\sec{i}})\,2\pi r\,dr
\end{equation}
where $i$ is the disk inclination angle (90\degr\ is edge-on), $d$ is the 
distance, $r_0$ and $R_d$ are the inner and outer disk edges, respectively, 
and $B_{\nu}(T_r)$ is the Planck function at the given radial temperature.  
This model is completely described by a set of 9 parameters, \{$i$, $r_0$, 
$R_d$, $\kappa_0$, $\beta$, $T_1$, $q$, $\Sigma_5$, $p$\}.

With only the unresolved SED, it is not possible to uniquely determine most of 
these parameters \citep[][see also Chiang et al.~2001 for a similar discussion 
with more sophisticated models]{thamm94}.  In light of this fact, the standard 
adopted procedure is to fix a subset of the parameters based on reasonable 
assumptions, usually \{$i$, $r_0$, $R_d$, $\kappa_0$, $\beta$, $p$\}, and fit 
to the data to determine the others \citep[e.g.,][]{bscg90,andrews05}.  
However, when information about the spatial distribution of emission at one or 
more frequencies becomes available, some of the inherent parameter degeneracies 
can be broken.  The flat disk model can be easily adapted to generate 
two-dimensional images at the expense of introducing the disk orientation, or 
position angle (PA), as an additional parameter. 

\subsection{Fitting Methodology and Data}
To estimate flat disk model parameters for this sample, we adopt a minimization 
technique that simultaneously fits both the SED and the spatial distribution of 
continuum emission at one submillimeter wavelength.  The latter is treated in 
the Fourier plane to avoid the nonlinearities associated with deconvolution and 
to properly account for the spatial response of the interferometer.  To further 
simplify and expedite the process computationally, we utilize the circular 
symmetry of the flat disk model (after accounting for inclination and 
orientation effects) to represent the spatial distribution of emission via the 
one-dimensional visibility profile, $V_{\nu}(s)$, the vector-averaged 
visibilities in annular bins of spatial frequency distance, $s = 
\sqrt{u^2+v^2}$.  To summarize, the minimization process is as follows: (1) for 
a given parameter set, generate a SED and submillimeter image as described in 
the previous section; (2) take the Fourier transform of the image and sample it 
only at the same spatial frequencies used in the observations; (3) bin the 
sparsely sampled Fourier transform of the image into a visibility profile; and 
(4) calculate the combined $\chi^2$ value for the SED and visibility profile.  
This minimization method is basically a hybrid of those developed by 
\citet{lay97} and \citet{kitamura02}.  The significant differences from the 
former \citep[see also][]{akeson98,akeson02} are that we choose to 
simultaneously use SEDs in the fits and to use vector-averaging for the 
visibilities rather than scalar-averaging (for the visibility amplitudes).  The 
main departure from the method adopted by \citet{kitamura02} is our preference 
to use the visibilities rather than synthesized images in the fits.  
 
Data from the literature were used to compile full SEDs for the survey sample,
with references given in the Appendix for individual disks.  Because 
near-infrared and shorter wavelengths are sensitive to emission from the 
extincted stellar photosphere, accretion, and the inner disk rim, only 
wavelengths of $\sim$8\,$\mu$m or longer were used in the fits.  Each SED was 
de-reddened based on the $A_V$ values listed in Table \ref{stars_table} and the 
interstellar extinction law compiled by \citet{mathis90}.  The exact extinction 
values are insignificant at the fitted wavelengths unless $A_V$ is very high.  
When available, \emph{IRAS} flux densities were color-corrected following the 
prescription of \citet{beichman88}.  Errors on the flux densities were computed 
as the quadrature sum of statistical and systematic uncertainties, the latter 
based on the uncertainties in the absolute calibration scales.  Continuum 
visibility profiles were generated from the SMA data with a typical bin width 
of 15\,k$\lambda$.  The visibility profile errors represent both the 
(comparatively small) statistical error on the average and the standard 
deviation in each bin.

\subsection{Parameter Estimation}
A properly sampled 10-dimensional parameter grid for this minimization 
technique would be computationally prohibitive and unwarranted given the 
typical signal-to-noise ratios in the data.  Fortunately, some of these 
parameters can be estimated independently.  We have chosen to fix values of 
\{$i$, PA, $\kappa_0$, $\beta$, $r_0$\} for each individual disk.  The opacity 
spectrum is discussed in detail in a later section, but here it is defined so 
that $\beta = 1$ and $\kappa_0 = 0.1$\,cm$^2$ g$^{-1}$ at 1000\,GHz \citep[this 
value implicitly assumes a 100:1 mass ratio between gas and dust;][]{bscg90}.  
Long-baseline near-infrared interferometry \citep{akeson05,eisner05} and models 
of the infrared SED \citep{muzerolle03} can provide independent estimates of 
the inner radius and inclination.  When these are not available we fix $r_0 = 
0.1$\,AU as a typical value, knowing that this parameter does not significantly 
impact either the SED or the submillimeter visibility profile (provided it is 
not too large).  Inclinations for some disks in the sample have been inferred 
from the kinematics of molecular line emission \citep{simon00} or scattered 
light signatures \citep{bouvier99}.  In other cases, we have estimated $i$ from 
elliptical Gaussian fits to the continuum visibilities (see Table 
\ref{cont_table} and discussion in \S 3.4).  Orientations have also been 
derived from the elliptical Gaussian fits, with supporting constraints based on 
the CO emission if available.

\begin{figure}
\epsscale{1.0}
\plotone{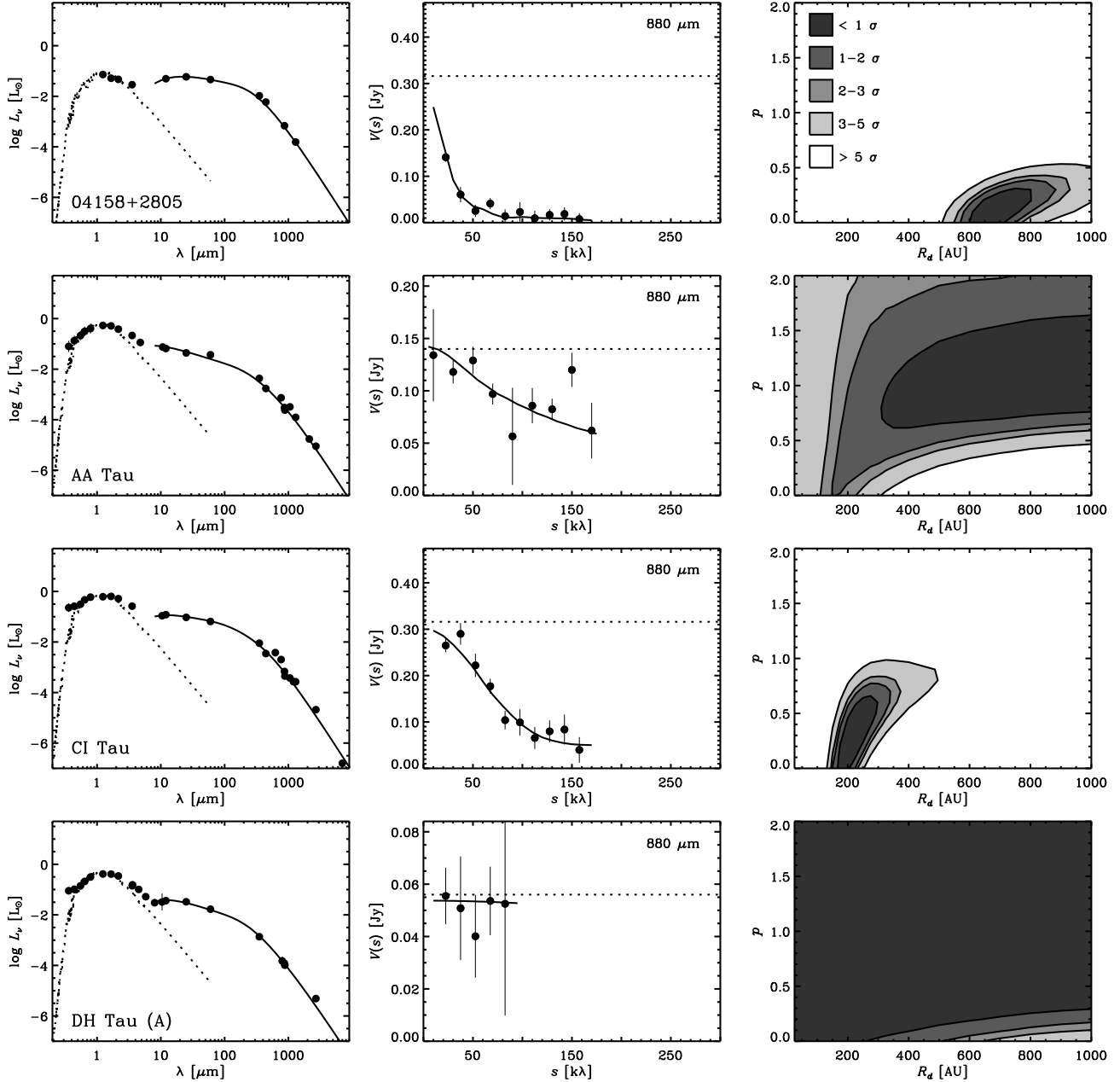}
\vspace{0.2cm}
\figcaption{(\emph{left}): De-reddened SEDs (defined so that $L_{\nu} = 4 \pi 
d^2 \nu F_{\nu}$ in solar luminosity units).  Dotted curves show 
\citet{kurucz93} models of the stellar photosphere.  (\emph{middle}): 
Visibility profiles.  The dotted line indicates the single-dish flux density 
scaled to the observing wavelength (labeled in the upper right corner for each 
disk) using a power-law fit to the submillimeter SED (see \S 4.2).  
(\emph{right}): Maps of $\Delta \chi^2$ from the minimizations based on the 
flat disk model, projected into the \{$R_d$, $p$\}-plane.  Contours and 
grayscale represent confidence intervals as indicated by the key in the top 
right panel.  The best-fit models are overlaid on the SEDs and visibility 
profiles, generated with the parameters given in Table \ref{structure_table}. 
\label{data_models1}}
\end{figure}

The remaining 5-parameter set \{$T_1$, $q$, $\Sigma_5$, $p$, $R_d$\} is 
constrained by the data using $\chi^2$-minimization.  Figures 
\ref{data_models1}$-$\ref{data_models6} show the SED and visibility profile 
data with best-fit models overlaid, along with $\Delta \chi^2$ maps projected 
into \{$R_d$, $p$\}-space for the sample disks.\footnote{For clarity, the 
best-fit visibility profiles are overplotted assuming \emph{uniform} spatial 
frequency coverage.  The actual fitting, however, is conducted using the 
sparsely-sampled coverage dictated by the observations.}  The grayscale and 
contours in the $\Delta \chi^2$ maps represent confidence intervals ranging 
from $<1$\,$\sigma$ to $>5$\,$\sigma$ from dark to light, as indicated in the 
keys in Figures \ref{data_models1} and \ref{data_models6}.  The values of $p$ 
and $R_d$ were limited in the fitting to reasonable ranges; $p \in$ \{0.0, 
2.0\} and $R_d \in$ \{25, 1000\}.  The values of the other fitted parameters 
were determined by refining a search over a wide range to a smaller subset with 
a lower $\Delta \chi^2$ threshold \citep[this is the same as the method devised 
by][]{lay97}.  The resulting best-fit disk structure parameter values and their 
1\,$\sigma$ errors are compiled in Table \ref{structure_table}, as well as 
total disk mass estimates ($M_d$, from a direct integration of the best-fit 
surface density profile), and values and references for the fixed parameter 
subset \{$r_0$, $i$, PA\}.  The total reduced $\chi^2$ values for the sample 
range from 0.5 to $\sim$3, with a median value of $\sim$1.8.  In many cases, 
much of the discrepancy between the model and the data can be attributed to 
absolute flux calibration uncertainties in the SED, particularly at 
submillimeter wavelengths where calibration accuracy is especially difficult.

The data for the Class I object WL 20 (S) was excluded from these fits because 
the SED has the clear steep rise in the infrared expected from extended 
envelope emission (see Fig.~\ref{data_models6}).  Modifications to the 
minimization description above had to be made for 3 other disks in the sample.  
The DM Tau and GM Aur disks are known to have large central holes \citep[i.e., 
large $r_0$;][]{calvet05}, which significantly affect their infrared SEDs and 
therefore the ability to infer their temperature distributions.  In these 
cases, we have chosen only to model the submillimeter part of the SED ($\lambda 
\ge 350$\,$\mu$m) and the continuum visibilities by fixing the temperature 
distribution.  The adopted outer disk ($r \ge 30$\,AU) temperature 
distributions are based on power-law approximations of the midplane temperature 
distributions computed by \citet{dalessio05} from detailed models of irradiated 
viscous accretion disks.  The \citet{dalessio05} models were chosen to have 
central stars with roughly the same spectral types and ages ($\sim$1\,Myr) as 
these disks, along with appropriate mass accretion rates (\emph{\.{M}} $\approx 
10^{-9}$ and $10^{-8}$\,M$_{\odot}$ yr$^{-1}$ for DM Tau and GM Aur, 
respectively) and other standard disk properties ($\alpha = 0.01$, grain size 
distribution index of 3.5, maximum grain size of 1\,mm).  The same procedure 
was adopted for DoAr 25 (assuming an age of 1\,Myr and \emph{\.{M}} $\approx 
10^{-9}$\,M$_{\odot}$ yr$^{-1}$), due to the peculiar morphology of the 
infrared SED (see Fig.~\ref{data_models4}).  Although it may be more likely 
that the \emph{IRAS} photometry for this source is contaminated by extended 
cloud emission, we can also not rule out a central hole similar to GM Aur, 
especially given the limit placed on the mass accretion rate, \emph{\.{M}} 
$\lesssim 6 \times 10^{-10}$\,M$_{\odot}$ yr$^{-1}$ \citep{natta06a}.  

\subsection{Caveats and Parameter Relationships}
Although the flat disk model is a computationally expedient approximation of a 
more detailed structure model, it contains a number of implicit assumptions 
which deserve to be highlighted.  Of these, the most significant is the 
requirement that the spatial distributions of temperature and density follow 
power-law behaviors with the same index ($q$ and $p$, respectively) 
\emph{across the entire extent of the disk}.  Furthermore, the vertical 
structure of the disk is altogether ignored, despite the fact that flaring is 
expected based on hydrostatic equilibrium \citep{kenyon87,chiang97,chiang99}.  
An over-simplified opacity function, particularly its presumed spatial 
uniformity in the disk, represents another critical assumption in the model.  
In general, each of the individual parameters in the flat disk model can have a 
significant impact on both the SED and visibility profile.  

In a fit of the flat disk model to a dataset, the radial temperature profile 
(parameters \{$T_1$, $q$\}) is essentially fixed by the intensity and shape of 
the infrared SED.  This is due to the fact that for the standard opacity 
prescription and any reasonable density values, the infrared emission generated 
in the inner disk is completely optically thick.  In this case $(1 - 
e^{-\tau_{\nu,r}\sec{i}}) \approx 1$, and Equation 4 can be simplified so that
\begin{equation}
F_{\nu} \propto T_1^{2/q} \nu^{3-2/q} \cos{i}, 
\end{equation}
where the normalization basically consists of physical constants 
\citep[cf.,][]{bscg90}.  In essence, the infrared SED is a sensitive diagnostic 
of the temperature profile, and not density.  The parameters \{$T_1$, $q$\} 
\emph{can} be determined through the SED alone, provided there are enough 
infrared datapoints.\footnote{This is still assuming the inclination and inner 
radius are fixed.  The choice of \{$r_0$, $i$\} can substantially influence the 
inferred temperature profile.}  The important point is that the inner, 
optically thick part of the disk sets the temperature profile, which is assumed 
to extrapolate into the outer regions.

In contrast, the disk size and density profile (parameters \{$\Sigma_5$, $p$, 
$R_d$\}) wield significantly more influence over the emission at longer 
wavelengths.  Assuming the disk is not too dense and the usual opacity law 
applies, most of the submillimeter emission should be optically thin 
\citep{bscg90}.  In this case $(1 - e^{-\tau_{\nu,r}\sec{i}}) \approx 
\tau_{\nu,r}\sec{i} = \kappa_{\nu} \Sigma_r \sec{i}$, and the flux density from 
Equation 4 becomes
\begin{equation}
F_{\nu} \propto \kappa_{\nu} \int_{r_0}^{R_d} B_{\nu}(T_r) \Sigma_r\,r\,dr \propto \kappa_{\nu} \nu^2 T_1 \Sigma_5 \int_{r_0}^{R_d} r^{1-p-q}\,dr \sim \frac{\kappa_{\nu} \nu^2 T_1 \Sigma_5}{(2-p-q)} R_d^{2-p-q},
\end{equation}
where we have assumed that the the Rayleigh-Jeans limit applies and $R_d \gg 
r_0$.\footnote{This relation holds for $p+q \ne 2$.  If $p+q = 2$, the 
optically thin flux density goes as $F_{\nu} \propto \kappa_{\nu} \nu^2 T_1 
\Sigma_5 \ln{R_d}$.}  The integrand in the second proportionality gives the 
radial surface brightness, $I_{\nu}(r)$, and the visibility profile, 
$V_{\nu}(s)$, as its Fourier transform
\begin{equation}
I_{\nu}(r) \propto r^{1-p-q} \,\,\,\, \stackrel{\rm{FT}}{\longrightarrow} \,\,\,\, V_{\nu}(s) \propto s^{p+q-3},
\end{equation}
where $s = \sqrt{u^2+v^2}$ is the spatial frequency distance 
\citep[cf.,][]{looney03}.  Equations 6 and 7 are simplified representations of 
a more complex situation.  In reality some of the submillimeter emission is 
generated in the inner disk where optical depths are high, thereby complicating 
the parameter relationships \citep[for details, see][]{bscg90}.  However, these 
relations are good first-order approximations and serve to illustrate a key 
component of the flat disk model: the outer, more optically thin part of the 
disk sets the density profile, which is then assumed to extrapolate into the 
inner regions.  

To illustrate these dependencies with an example related to fitting real data, 
consider an adjustment that decreases the value of the power-law index of the 
surface density profile, $p$ (i.e., a redistribution of mass to larger radii).  
This would act to increase both the submillimeter flux densities and the 
surface brightness at larger radii, thus making the visibility profile drop off 
more rapidly at shorter spatial frequency distances (see Equations 6 and 7).  
Keeping in mind that the temperature profile is set in the infrared, 
compensating for the drop of $p$ in a fit would require a decrease in the 
surface density normalization ($\Sigma_5$) and/or outer radius ($R_d$).  This 
interplay between parameters is a manifestation of the submillimeter emission 
in the outer disk being mostly optically thin.  As would be expected, a sharp 
outer radius cutoff more strongly affects the observables for shallower density 
profiles (i.e., lower $p$), whereas the outer radius quickly becomes negligible 
(and thus poorly constrained) as the density profile steepens \citep[see 
also][]{mundy96}.  

The above discussion reiterates that the assumptions of single power-laws for 
the temperature and density across the entire disk are critical.  Without them, 
we could not assume that the inner disk temperature profile is applicable in 
the outer disk or vice versa for the outer disk density profile.  The other 
parameters that were fixed in the model have more complicated effects on the 
data.  To start, the location of the inner radius really affects only the 
infrared SED, such that larger $r_0$ would decrease fluxes at those 
wavelengths.  Generally this is not an issue, as the shorter infrared 
wavelengths are not used in the fits (however, see the Appendix regarding disks 
with large $r_0$).  The effects of the opacity can be significant, and are 
discussed in their own right in the following sections.  

The projected disk geometry, characterized by \{$i$, PA\} and fixed in the 
minimization algorithm used here, can significantly influence the other 
parameters in the flat disk model.  Several of the disks in the sample have 
reliable geometry constraints from independent measurements (see the notes in 
Table \ref{structure_table}), but in most cases we derive \{$i$, PA\} from 
elliptical Gaussian fits to the continuum visibilities.  In such cases, a 
natural concern is that these Gaussian geometries do not represent the true 
disk geometries.  To test the validity of using the Gaussian fits, we created 
synthetic continuum images using the flat disk formalism described above with a 
variety of parameter sets, inclinations, and orientations.  These synthetic
images were then ``observed" in the same way as the data, for both the compact 
(C2) and extended (E) SMA configurations and with the appropriate thermal noise 
added to reproduce typical signal-to-noise levels.  The geometries derived from 
elliptical Gaussian fits to the synthetic visibilities can then be compared to 
the known input geometries from the flat disk model.  In general, the Gaussian 
inclinations yield fairly accurate results, but a few important points should 
be addressed.  First, inclinations based on Gaussian fits are better 
approximations when surface brightness profiles are shallow (e.g., for lower 
values of $p$).  Second, coverage in the Fourier plane is important: the 
Gaussian inclinations are more accurate with the combined C2+E array 
configuration (note that the C1 configuration is basically intermediate to the 
C2 and C2+E cases).  And third, the Gaussian fits tend to slightly 
underestimate intermediate and high inclinations (by $\sim$10\degr\ at most), 
and overestimate low inclinations (by up to $\sim$15$-$20\degr).

To investigate the effects of fixing an erroneous value of the inclination in 
the fitting process, we also generated synthetic SEDs for the disk models 
described above.  We then fit the synthetic SEDs and visibilities for a range 
of fixed inclination angles, and the best-fit parameter values were compared 
with the known input values.  In general, fixing a value of $i$ that is lower 
than the true value leads to underestimates of other parameters, most 
importantly \{$\Sigma_5$, $p$, $R_d$\}.  The magnitude of this effect depends 
on several conditions: the true inclination and how severely the assumed 
inclination underestimates it; the radial surface brightness distribution 
(i.e., $p$); the relative projections of the surface brightness distribution 
and the Fourier coverage of the interferometric observations; and the 
signal-to-noise ratio.  As an example, consider a flat disk with a typical 
inclination of 60\degr.  Given the underestimate of $i$ expected from a 
Gaussian fit, our simulations indicate that we could underestimate $p$ by up to 
$\sim$0.2$-$0.3 if the true value is $p \approx 1.5$.  The underestimate of $p$ 
is insignificant if $p \lesssim 1$ (although the other parameters are 
changed).  In the end, we are forced to adopt Gaussian inclinations as the best 
information available for many of the sources.  In general, we are unlikely to 
overestimate the inclination in this way simply because nearly face-on disks 
should be relatively rare (assuming disk geometries are random).  The sense of 
any effects on the parameters is then as described above, although the exact
magnitude is generally unknown.  

The analysis described above was also applied to better understand how fixing 
the disk orientation (PA) can affect the fitting results.  Fortunately, the 
Gaussian fits effectively reproduce the true orientations in most cases within 
the errors (see Table \ref{cont_table}).  The exceptions are for low 
inclination angles (where the PA is basically meaningless anyway) or small 
radii.  Again, any effects the PA could have on the visibility profile depend 
on the relative orientations of the projected disk emission and the 
sparsely-sampled Fourier coverage afforded by the observations.  However, the 
agreement between the Gaussian fits and simulated data gives us confidence that 
fixing the disk orientations does not significantly affect the fitting 
results.

\section{Results}
\subsection{Disk Structure}
Keeping in mind the caveats discussed above, the combined fits of SEDs and 
visibilities provide constraints on basic structure parameters for the large 
number of circumstellar disks in this sample.  Despite the simplicity of the 
adopted model and the remaining parameter uncertainties due to limited 
sensitivity and spatial resolution, we find the following: (\emph{a}) the 
variation of temperature with radius is intermediate to the idealized cases of 
flat and flared disks; (\emph{b}) small outer radii can be ruled out for most 
cases; and (\emph{c}) the density apparently drops slowly with radius in the 
outer disk.

\subsubsection{Disk Temperature Profiles}
Figure \ref{Tr_dists} shows the distributions for the radial temperature 
profile parameters $T_1$ and $q$ for the sample.  These histograms were created 
to crudely account for the uncertainties in the parameter measurements as 
follows.  For each disk, we generate a Gaussian distribution normalized to have 
an area of 1 with a mean and standard deviation corresponding to the best-fit 
parameter value and error in Table \ref{structure_table}.  Parameters with 
asymmetric errors are allowed to have different standard deviations with a 
discontinuity at the mean (to ensure a total area of 1 for the distribution).  
All of the individual error distributions are then summed and binned (which 
effectively smooths away any discontinuities).  In this way, the contribution 
of each disk to the histogram is appropriately weighted according to the 
uncertainty in the parameter measurement \citep[e.g.,][see their 
Fig.~10]{clayton04}.  The median values for each parameter are shown with 
dotted vertical lines.  The 1\,AU temperatures in this sample have a clear, but 
fairly broad peak centered at $\sim$200\,K.  

The power-law indices (i.e., the radial slope of the temperature profile) also 
have a peak around the median value $q \approx 0.62$, with a few sources having 
substantially shallower temperature distributions (lower $q$).  This peak in 
the distribution of the power-law index falls between the idealized values for 
the flat case, $q \approx 0.75$ \citep{adams87}, and the flared case, $q 
\approx 0.43$ \citep[e.g.,][]{chiang97}.  This is probably the result of 
imposing a simple model upon a more complicated reality.  For a typical disk, 
the temperature distribution is determined by the infrared SED from 
$\sim$8$-$60\,$\mu$m, corresponding to emission produced roughly between 
$\sim$0.5$-$30\,AU (give or take a factor of 2).  In detailed physical models 
of disk structure, this corresponds to the region where vertical flaring begins 
to affect the temperature distribution; in essence, this is the transition 
region between the flat and flared scenarios 
\citep[e.g.,][]{dalessio98,dalessio99}.  It is therefore not so surprising that 
the typical \emph{single} power-law index derived here lies between the 
\emph{two} indices which better describe a more realistic disk structure.

This point is illustrated in Figure \ref{Tr}, which serves as a comparison of 
the typical temperature profiles measured here and those for more sophisticated 
models which self-consistently treat a variety of heating mechanisms in a more 
realistic two-dimensional disk structure 
\citep{dalessio98,dalessio99,dalessio05}.  The datapoints in this figure mark 
the median temperatures for the sample at various radii, determined based on 
the best-fit $T_r$ parameters listed in Table \ref{structure_table}.  The error 
bars show the first and third quartile temperatures to represent the range in 
the sample.  The midplane radial temperature distribution for a detailed model 
of an irradiated accretion disk is also shown for comparison 
\citep[cf.,][]{dalessio05}.  The example model assumes a stellar mass of 
0.5\,M$_{\odot}$, effective temperature of 4000\,K, and age of 1\,Myr (all 
median values for the sample), and standard disk parameters \citep[\emph{\.{M}} 
= $10^{-8}$\,M$_{\odot}$ yr$^{-1}$, $\alpha = 0.01$, grain size distribution 
index of 3.5, and maximum grain size of 1\,mm; see the descriptions 
in][]{dalessio05}.  In general, the simplified temperature distributions 
derived here turn out to be reasonable approximations of those determined with 
a more sophisticated treatment.  However, the effects of the discrepancies 
between the simple approximations and the detailed models will be revisited 
below.

\subsubsection{Disk Density Profiles}
The distributions of the radial surface density profile parameters $\Sigma_5$ 
and $p$ are shown in Figure \ref{SDr_dists}, created in the same way as for 
those in Figure \ref{Tr_dists}.  The distribution of $p$ should be treated with 
particular caution, as the errors on individual measurements are large.  The 
disks with completely unconstrained $p$ values are not included in the 
distributions in Figure \ref{SDr_dists} or the analysis that follows.  The 
distribution of $\Sigma_5$ has a broad peak around a median value of $\sim$14 
g cm$^{-2}$.  The distribution of $p$ indicates more disks in this sample have 
low values (i.e., less than $\sim$1) than high, with a median value of $p 
\approx 0.5$.  The standard of reference often used in discussing the density 
distribution in circumstellar disks is the minimum mass solar nebula (MMSN), 
the progenitor circumstellar disk around the Sun.  The MMSN density 
distribution is constructed by augmenting the masses of the planets in the 
solar system until cosmic abundance values are obtained, and then smearing the 
mass out into annuli centered on each planet's semimajor axis 
\citep{weidenschilling77}.  A simple power-law fit to the MMSN surface density 
distribution indicates $\Sigma_5 \approx 150$\,g cm$^{-2}$ and $p \approx 1.5$ 
\citep{hayashi85}, both of which are marked in Figure \ref{SDr_dists}.  As a 
second reference case, a steady-state viscous accretion disk has $p \approx 1$, 
while the normalization depends on a variety of other parameters 
\citep[e.g.,][]{hartmann98}.

The best-fit values of $p$ for this sample, while not very strongly 
constrained, often indicate that the density drops off more slowly in the outer 
disk than is expected from either the MMSN or the standard viscous accretion 
disk models.  This is illustrated in Figure \ref{SDr}, where again a comparison 
can be made between the sample median density distribution and a more 
sophisticated disk structure model.  As with Figure \ref{Tr}, we plot sample 
median surface densities at some representative radii (determined from the 
best-values of \{$\Sigma_5$, $p$\} in Table \ref{structure_table}), with error 
bars marking the first and third quartiles as an indication of the range in the 
sample.  Overlaid on the plot as a solid curve is the density distribution for 
the same irradiated accretion disk model described above for Figure \ref{Tr} 
\citep[cf.,][]{dalessio05}.  The MMSN density distribution is also plotted as a 
dashed line.  Observational measurements of the value of $p$ in circumstellar 
disks are critical to developing a better understanding of disk evolution and 
the planet formation process.  Therefore, an examination of possible causes for 
the apparent disagreement between the typical values measured here ($p 
\lesssim$ 1 in most cases) and those expected from theory ($p \gtrsim 1$) is 
necessary.

In \S 3.4, we already highlighted a possible cause of artificially low values 
of $p$ in this sample; namely, underestimates of the true disk inclination 
angles.  To examine this issue in more detail, we look at the cases where 
independent inclination estimates are available (see col.~11 in Table 
\ref{structure_table}).  For these disks, the median best-fit value of $p$ 
remains the same as for the full sample, $\sim$0.5.  Therefore, for at least 
some of the disks in this sample, the low measured values of $p$ do not arise 
from erroneous inclination values.  Without individual constraints on $i$ for 
the remaining sample disks, we are forced to rely on the Gaussian fits.  As 
discussed in \S 3.4, any induced decrease in $p$ is typically expected to be 
small ($\sim0.2-0.3$ if $p > 1$).  

Another potential issue with the surface density constraints could arise due to 
the simplified treatment of the disk temperature distribution.  As mentioned 
above, detailed disk models indicate that the temperatures in the outer disk 
have a relatively shallow dependence on radius (due to vertical flaring) 
compared to the inner disk, where the temperature distribution is actually 
constrained by the data \citep{chiang97,dalessio98,dalessio99}.  Because of the 
single power-law assumption for $T_r$ made here, the derived power-law indices 
($q$) are usually at intermediate values to the flat and flaring regimes in a 
more realistic disk.  This implies that our temperature distributions could be 
too steep in the outer regions; the values of $q$ may be too high.  If we were 
to adopt a lower value of $q$ more consistent with the detailed models, 
Equation 7 indicates that the value of $p$ should be \emph{increased} in order 
to maintain the shape of the visibility profile.  Therefore, the simplified 
temperature description used here could lead to underestimates of the surface 
density power-law index, $p$.  In essence, since the quantity $p+q$ is 
well-constrained by the visibility profile, $\Delta p \approx - \Delta q$.  
This should only lead to modest increases on the order of $\Delta p \approx 
0.2-0.3$, corresponding to the expected difference between the typical measured 
$q$ values and the flared disk $q$ values.  

Other simplified assumptions in the model are not expected to significantly 
affect our determination of the power-law index of the density distribution.  
One such example is the assumed sharp outer disk boundary, compared to the 
exponentially decreasing density profile expected from the similarity solutions 
for viscous accretion disks \citep{lyndenbell74,hartmann98}.  Using the 
solutions described by \citet{hartmann98}, we generated synthetic SEDs and 
visibilities with reasonable errors and modeled them with the procedure 
described in \S 3.  As one might expect, these simulations generally reproduce 
the correct value of $p$, or even slightly higher values due to the steep 
exponential drop-off in the outer disk.  Therefore, the assumption of a sharp 
outer boundary does not factor into the low $p$ values inferred for much of our 
sample.  In the end, there are two compelling reasons that may explain the 
shallow outer disk density distributions: over-simplified temperature 
distributions and possible inclination underestimates.  The correction for the 
former is expected to be an additive increase $\Delta p \approx 0.2-0.3$.  The 
magnitude of the adjustment to $p$ for the latter is more difficult to 
predict.  Assuming that most of the disks in this sample have an intermediate 
inclination, we estimate that the correction is typically $\Delta p \approx 
0.1$.  Given the rather poor constraints on $p$ in general and these implied 
corrections, we suggest that the typical disk in this sample has $p \approx 
0.7-1.0$.  

\subsubsection{Disk Sizes and Masses}
In addition to the density structure, the spatially resolved SMA measurements 
provide valuable new constraints on circumstellar disk sizes.  The top panel of 
Figure \ref{RM_dist} shows the distribution of the outer disk radii for this 
sample, constructed in the same way described above for the temperature and 
density profile parameters.  The distribution has a peak near the median radius 
of $\sim$200\,AU, with a roughly Gaussian shape and extended large radius 
wing.  Most of the disks are at least partially resolved, with radii larger 
than $\sim$100\,AU.  The values of $R_d$ measured here technically represent 
some characteristic radius beyond which the temperature/density conditions in 
the disk do not produce substantial submillimeter emission compared to the 
noise levels in the data.  True disk sizes could be larger, but would have to 
be measured with more optically thick tracers like molecular line emission 
\citep[e.g.,][]{simon00} or optical silhouettes 
\citep[e.g.,][]{mccaughrean96}.  Finally, as a caution it should be emphasized 
that elliptical Gaussian fits to interferometer data tend to 
\emph{underestimate} disk sizes (compare the FWHM sizes in Table 
\ref{cont_table} with the $R_d$ values in Table \ref{structure_table}).  The
problem is really one of contrast, so that the level of discrepancy increases
dramatically for more centrally-concentrated surface brightness distributions
(e.g., for higher values of $p$).

Armed with constraints on both the density distribution and size of the disk, 
the total mass is computed by simply integrating the surface density over the 
disk area.  The distribution of disk masses for the sample is shown in the 
bottom panel of Figure \ref{RM_dist}.  A broad peak is seen around the median 
mass of 0.06\,M$_{\odot}$.  It is important to emphasize again that the sample 
selection criteria explicitly introduce a bias toward more massive disks, and 
so this distribution in particular is not representative of the typical T Tauri 
disk \citep{andrews05}.  Despite the rather large uncertainties in the density 
distribution and radius parameters, the inferred disk masses are constrained 
within a factor of 2.  This is due to the low optical depths where the density 
and radius parameters are measured.  Larger uncertainties in the disk masses 
are due to our limited knowledge of the opacity and the density structure in 
the inner, optically thick regions of these disks (both unaccounted for in 
Table \ref{structure_table} and Figure \ref{RM_dist}; see \S 5). 

No statistically significant correlations between the best-fit disk structure 
parameters and properties of the central stars (e.g., $T_*$, $M_*$, age) or 
other disk diagnostics (e.g., \emph{\.{M}}) were found for this sample.  Only a 
marginal trend of increasing 1\,AU temperatures for earlier spectral types is 
noteworthy.  Neither are there any significant differences between disk 
structure properties among single and multiple star systems, or the two 
different star-forming regions incorporated into the sample.  This is perhaps 
not so surprising, considering the limited ranges of star/disk properties in a 
sample of this size and the remaining uncertainties on the structure 
parameters.  

\subsection{Opacity}
Perhaps the greatest uncertainties in studies of circumstellar disks are 
related to the growth of grains into larger solids and how this process 
subsequently shapes the structure, dynamics, and evolution of disk material 
\citep[see][]{beckwith00,dominik06}.  In the high density environment of a 
circumstellar disk, dust grains are expected to grow via collisional 
agglomeration and gravitationally settle toward the midplane of the disk.  
Observational evidence for the combined effects of grain growth and 
sedimentation have been accumulating from a variety of techniques \citep[see 
the review by][]{natta06}.  At submillimeter wavelengths, the growth of solids 
in the disk is manifested as a change in the opacity spectrum, and therefore 
the shape of the continuum SED.  Increased grain growth leads to grayer 
opacities and more efficient submillimeter emission \citep[i.e., a shallower 
SED slope; e.g.,][]{beckwith91,calvet02,natta04}.  The key affected parameter 
is the power-law index of the opacity spectrum, $\beta$, which is expected to 
\emph{decrease} as a result of the grain growth process \citep{miyake93,pollack94,krugel94,henning95,henning96,dalessio01,dalessio06,draine06}.

For an optically thin disk in the Rayleigh-Jeans limit, Equation 6 shows that 
the submillimeter SED has a power-law behavior with frequency, $F_{\nu} \propto 
\kappa_{\nu} \nu^2 \propto \nu^{2+\beta}$.  In this case, one could simply fit 
the observed SED with a general form $F_{\nu} \propto \nu^{n}$ and determine 
the opacity index, $\beta = n-2$, directly.  However, the submillimeter 
emission may not be completely optically thin.  Some optically thick emission, 
for which $F_{\nu} \propto \nu^2$, would serve to decrease the measured 
quantity $n$ and result in misleadingly low values of $\beta$.  Therefore, the 
ratio of optically thick to thin emission, $\Delta$, needs to be explicitly 
calculated to translate the measured submillimeter SED power-law index, $n$, 
into the opacity index, $\beta$, via the simple approximation $\beta \approx 
(n-2)(1 + \Delta)$ \citep[cf.,][]{bscg90,beckwith91}.  The ratio $\Delta$ 
depends on wavelength and the disk structure (particularly \{$q$, $\Sigma_5$, 
$p$, $R_d$, $i$\}).  Multiwavelength submillimeter photometry from single-dish 
telescopes and reasonable assumptions for $\Delta$ indicate that $\beta \approx 
1$ for a large number of circumstellar disks 
\citep{weintraub89,beckwith91,mannings94,andrews05}.  This is significantly 
lower than the value for the interstellar medium, where $\beta_{\rm{ISM}} 
\approx 1.7$ \citep{hildebrand83,weingartner01}, indicating an evolution in the 
opacity spectrum that could be due to the growth of grains.  High-resolution 
interferometric work has allowed measurements of $\Delta$ and therefore less 
ambiguous constraints on $\beta$, adding further evidence for the presence of 
large grains in a variety of disks 
\citep[e.g.,][]{testi01,testi03,natta04,wilner05,rodmann06}.

Following the method described above, Table \ref{opacity_table} lists the 
measured values of $n$, $\Delta$, and $\beta$ for the sample disks.  The SED 
power-law indices, $n$, were determined using only wavelengths longer than 
850\,$\mu$m (or 1\,mm whenever possible) to ensure that the Rayleigh-Jeans 
criterion holds.  Based on the temperatures at the outer disk edges, any 
corrections for violating this criterion would lead to only modest increases in 
$n$ (generally within the 1\,$\sigma$ error bars).  For each disk, the ratio of 
optically thick to thin emission, $\Delta$, at the shortest wavelength used in 
determining $n$ was computed for each parameter combination corresponding to 
the regions in Figures \ref{data_models1}$-$\ref{data_models6} within the 
1\,$\sigma$ confidence interval.  Listed in Table \ref{opacity_table} are the 
\emph{maximum} $\Delta$ values inferred from these regions (usually those for 
the highest $p$ and smallest $R_d$).  Figure \ref{beta_dist} shows the 
distribution of $\beta$ values calculated for this sample, created in the same 
way as other distributions presented here to account for the large error bars.  
The median value is $\beta \approx 0.7$, considerably lower than for the 
interstellar medium, and consistent with the typical estimates for 
circumstellar disks.  

However, this indirect method of deriving $\beta$ from the submillimeter SED 
and disk structure models contains some circular logic.  For a given disk, the 
value of $\Delta$ depends on the best-fit structure parameters, which were 
derived based on an \emph{assumed opacity spectrum}.  As an experiment to 
determine the influence of $\beta$ on the disk structure parameters, we fitted 
the data for a typical disk (WaOph 6) with a range of $\beta$ values.  The 
surface density normalization, $\Sigma_5$, proves to be the most dramatically 
affected parameter due to changes in $\beta$.  For the mostly optically thin 
submillimeter emission, the optical depth ($\tau_{\nu, r}$) is well-determined 
by the flux, and therefore $\Sigma_r \propto \kappa_{\nu}^{-1}$.  So, for a 
fixed $\kappa_0$ an increase in $\beta$ leads to a drop in $\kappa_{\nu}$, and 
subsequently an increase in $\Sigma_5$ to compensate for the observed 
emission.  As long as the emission at infrared wavelengths remains optically 
thick, the temperature distribution parameters are not significantly affected.  
For larger variations of $\beta$ (e.g., $\pm 1$), there can also be substantial 
changes in \{$p$, $R_d$\}, such that a larger $\beta$ leads to a larger radius 
and a smaller value of $p$.  As an aside in the context of the discussion in \S 
4.1.2, note how this implies that our assumption of a $\beta$ value lower than 
in the interstellar medium is also not responsible for artificially decreasing 
the measured $p$ values.

The above simulations imply that the optically thick area of the disk does not 
change significantly for different input $\beta$ values, and therefore the 
value of $\Delta$ remains roughly the same.  Given this tendency for 
self-correction, the method of determining $\beta$ by adjusting the measured 
SED slope, $n$, to accomodate an optically thick contribution is a reasonable 
approximation.  However, the smaller changes in \{$R_d$, $p$\} noted for 
different assumed $\beta$ values do have the capability to affect $\Delta$.  
Because of the interplay between these parameters, a better way to determine 
$\beta$ is by allowing it to be a free parameter in the fits, a task usually 
accomplished by fixing $p$ instead.  Given the data in hand, the addition of 
$\beta$ as another free parameter in the fits described here unfortunately does 
not provide any statistically meaningful constraints (without fixing other 
currently free parameters).  However, this could be accomplished either with 
multiwavelength interferometric data \citep[e.g.,][]{lay97} or data with very 
high quality and resolution \citep[see][]{hamidouche06}.

As Table \ref{opacity_table} indicates, we can not provide reliable estimates 
of $\beta$ for a number of disks in this sample for two primary reasons.  First 
are the cases where the submillimeter emission is either not spatially resolved 
(e.g., FT Tau) or highly optically thick due to the disk inclination (e.g., RY 
Tau), and therefore $\Delta$ (and thus $\beta$) could be arbitrarily high.  For 
these disks we quote only lower limits on $\beta$ based on the assumption that 
$\Delta \ge 0$.  And second are the cases where the observed power-law index of 
the submillimeter SED, $n$, is less than 2.  The submillimeter SED can have $n 
< 2$ for a variety of reasons, including low disk temperatures (i.e., the 
Rayleigh-Jeans criterion fails), extensive dust sedimentation 
\citep{dalessio06}, simple statistical errors, or contamination from extended 
and/or non-disk emission (e.g., non-thermal emission from a stellar wind).  
Unfortunately, the latter possibility of extended emission contamination, 
perhaps from residual envelopes surrounding the star/disk systems, seems to be 
especially likely for the more embedded sources in the Ophiuchus star-forming 
region.  A more reliable measurement of $n$ for these sources will require 
interferometric observations at a second wavelength to ensure that only compact 
disk emission is included in the SED fits.

\section{Discussion}
Including those presented here, roughly 40 circumstellar disks around T Tauri 
stars have been observed with (sub)millimeter interferometers at a variety of 
wavelengths and resolutions.  Nearly half of the 24 disks in this survey have 
been previously observed \citep[AA Tau, CI Tau, DL Tau, DM Tau, DN Tau, DR Tau, 
FT Tau, GM Aur, RY Tau, AS 209, and WL 20;][]{koerner93,koerner95,dutrey96,dutrey98,guilloteau98,looney00,simon00,barsony02,kitamura02,dartois03,rodmann06}.  
An unbiased comparison of disk structure constraints in the literature would be 
a nearly impossible task, due to the wide diversity of data and models adopted 
for different disks.  Nevertheless, it is worthwhile to summarize what has been 
observationally inferred about disk structure from the perspective of high 
resolution submillimeter measurements.  

There are relatively few constraints on the density distribution in these 
disks.  This is partly due to limited sensitivity and resolution, but also due 
to a common preference to constrain other parameters instead (e.g., \{$i$, 
$\beta$\}).  In that case, $p$ is usually fixed in the modeling to ensure a 
reasonable number of degrees of freedom and useful measurements of the 
interesting parameters.  When not fixed, a wide range of $p$ values have been 
noted.  One method to determine disk structure utilizes spectral images of CO 
rotational transitions and the submillimeter portion of the SED with a model 
similar to that described here, but including a flared vertical structure 
\citep{guilloteau98}.  In most cases this technique restricts the value of $p$, 
but the results provide good fits to the data for values near 1.5 
\citep{dutrey98,guilloteau98,dartois03,dutrey03}.  Other methods have more in 
common with those used here, based on both high resolution submillimeter 
continuum data and the SED.  These lead to a considerable range of $p$ values, 
from 0$-$1.5, depending on the individual disk and which data are used in the 
modeling \citep{mundy96,lay97,akeson98,akeson02,kitamura02,duchene03}.  The 
only other large collection of $p$ measurements to date is from the 2\,mm 
survey of \citet{kitamura02}; their work shows a median $p \approx 0.6$, 
similar to the value presented here.  If we take at face value all of the 
available measurements of $p$ determined with the flat disk model, the 
distribution is similar to that presented in Figure \ref{SDr_dists}, again with 
a median value of $\sim$0.5.  Upon consideration of the potential systematic 
underestimates of $p$ for these simple models (see \S 4.1.2), the typical disk 
probably has a value closer to $p \approx 0.7-1.0$.

In terms of the outer disk radius, the silhouette disks and proplyds in the 
Orion nebula provide an interesting comparison to the SMA sample.  
\citet{vicente05} compute disk sizes for 149 such objects in a homogeneous 
way.  Their resulting distribution of $R_d$ has a similar shape to that seen in 
the top panel of Figure \ref{RM_dist}, but the peak is substantially shifted to 
smaller radii.  The median proplyd disk radius is $\sim$70\,AU, while the 
subsample of silhouette disks has a median radius of $\sim$135\,AU.  A 
Kolmogoroff-Smirnov test confirms that the Orion disks are significantly 
smaller than the disks in Taurus and Ophiuchus, where the median radius is 
$\sim$200\,AU.  This difference is not likely to be an artifact of how the 
radii are measured.  The characteristic radii inferred from optical 
observations are expected to be systematically \emph{larger} than those at 
submillimeter wavelengths because the much higher optical depths at shorter 
wavelengths allow material to be traced out to lower densities (and thus larger 
radii).  The reason for the different disk sizes in Orion is most likely 
related to the local environment, perhaps due to dynamical interactions in a 
higher stellar density cluster and/or the intense external photoevaporation 
from nearby massive stars \citep[e.g.,][]{johnstone98}.  The disks in the 
Taurus and Ophiuchus star-forming regions do not have to contend with such 
extreme environments, and therefore have larger sizes than those in Orion 
despite their similar ages.

\subsection{Accretion Disks}
The values of $p$ inferred above are similar to the expectations for 
steady-state viscous accretion disks 
\citep{hartmann98,dalessio98,dalessio99,calvet02}.  This is also in good 
agreement with the $p \approx 1$ inferred for the carefully studied TW Hya disk 
in the context of those models \citep{wilner00}.  In the standard similarity 
solutions for the structure of an accretion disk, the surface density power-law 
index ($p$) corresponds to the power-law index ($\gamma$) for the radial 
distribution of the viscosity \citep[e.g.,][]{lyndenbell74,hartmann98}.  An 
independent constraint on $\gamma$ can be obtained via measurements of the 
decay of mass accretion rates with time, in essence of $\eta$ in the scaling 
relation \emph{\.{M}} $\propto t^{-\eta}$ \citep[][see their 
Eqn.~28]{hartmann98}.  Using the accretion rates of young stars in the Taurus 
and Chameleon star formation regions, \citet{hartmann98} inferred that $\eta 
\approx 1.5-2.8$, or equivalently $\gamma \approx 1.0-1.7$, with a preference 
for the lower value based on an assessment of the more likely systematic 
errors.  While these measurements of \{$p$, $\eta$\} have fairly large 
uncertainties, both are consistent with the viscosity being distributed with a 
roughly linear dependence on radius in the outer parts of circumstellar disks.

In such a case, the $\alpha$ parameter for the viscosity 
\citep{shakura73,pringle81} can be considered constant over the timescales of 
interest for these disks \citep{hartmann98}.  The $\alpha$ parameter 
essentially characterizes the level of turbulent viscosity in the disk, and 
therefore the rate at which the disk structure evolves; the sense is that lower 
values of $\alpha$ (i.e., less turbulent viscosity) correspond to slower 
evolution.  Angular momentum conservation leads to the expansion of accretion 
disks to larger radii, where the rate of expansion basically depends upon 
$\alpha$, the mass of the central object ($M_*$), and the initial disk 
conditions.  Assuming these initial conditions are roughly similar for T Tauri 
disks, the observed time variation of disk properties can be useful tools to 
estimate the value of $\alpha$.  Disk sizes are of particular interest in this 
case.  

However, the meaning of the sharp boundary $R_d$ measured in this paper is 
unclear in the context of the accretion disk models, where instead the density 
distribution is allowed to drop off exponentially.  The values measured here 
really correspond to a sensitivity limit to the surface brightness profile of 
the continuum emission.  To empirically relate our values of $R_d$ with some 
characteristic radius in the accretion disk formalism described by 
\citet{hartmann98}, we have used the same fitting scheme described in \S 3 to 
model simulated accretion disk SEDs and visibilities at various evolutionary 
stages.  The results for a fiducial model (see below) generally indicate that 
$R_d$ corresponds to an accretion disk radius that encircles a large fraction 
$f$ of the total disk mass \citep[cf.,][their Eqn.~34]{hartmann98}.  For 
$\alpha = 0.01$, the value of $f$ decreases smoothly from nearly 1 at 
$10^5$\,yr to $\sim$0.5 at $10^7$\,yr.  When $\alpha = 0.001$, $f \approx 1$ 
before $10^6$\,yr, where it begins to drop to $\sim$0.7 by 10$^7$\,yr.  

Figure \ref{evolve} shows plots of 850\,$\mu$m flux densities, mass accretion 
rates, disk radii, and Gaussian FWHM sizes as a function of stellar age for 
this sample.  Values of \emph{\.{M}} and ages were gathered from the 
literature, and are listed for individual disks in Table \ref{stars_table}.  
Overlaid on these plots are the expected evolutionary behaviors of these disk 
properties based on the $\gamma = 1$ similarity solutions for viscous accretion 
disks of \citet[][cf.~their \S 4.3.1]{hartmann98} for a fiducial parameter 
set.  Here we have fixed the initial disk mass to 0.1\,M$_{\odot}$, the stellar 
mass to 0.5\,M$_{\odot}$, the initial scaling radius to $R_1 = 10$\,AU, $d = 
150$\,pc (the average between the two regions used here) and have let $T_r 
\propto r^{-0.5}$ with a normalization of 200\,K at 1\,AU (the median value 
found here).  These models were computed for two values of $\alpha$; 0.01 
(solid curves) and 0.001 (dashed curves).  The 850\,$\mu$m flux densities for 
these accretion disk models were calculated using our Equation 4, with the 
surface density profile described by \citet[][their Eqn.~33]{hartmann98} and 
enforcing a minimum disk temperature of 7\,K.  The evolutionary behavior of the 
mass accretion rates shown in Figure \ref{evolve} were computed at $r = 0$, 
following \citet[][their Eqn.~35]{hartmann98}.  

The variation of the disk radius with time, shown in the lower left panel of 
Figure \ref{evolve}, corresponds to the accretion disk model prediction $R_d 
\propto R_1 \mathcal{T}$, where $R_1$ is a scaling radius 
(encircling $\sim$60\% of the disk mass at $t = 0$), $\mathcal{T}$ is a scaling 
time related to the number of viscous timescales that have passed, and the 
constant of proportionality is determined by the empirically measured $f$ value 
described above.  The time variation of the Gaussian FWHM size for these 
accretion disk models were determined by ``observing" synthetic images with the 
SMA C1 array configuration and fitting directly to the visibilities.  The 
datapoints plotted in this panel are FWHM values determined only from compact 
SMA configurations for ease of comparison (there is no significant difference 
between C1 and C2 for this purpose).  Different model curves and datapoints are 
shown for 880 and 1330\,$\mu$m.  Note that the evolutionary pattern of the 
model FWHM sizes here and that shown by \citet[][their Fig.~4]{hartmann98} are 
different.  This is due to our use of a fit to the visibilities, while they 
chose to convolve the image with a synthesized beam-sized Gaussian.  If their 
technique is adopted, we do reproduce the FWHM behavior shown in that paper.

For a more direct comparison of the observable data with the accretion disk 
model predictions, Figure \ref{median_data} shows a sample median SED and 
median visibility profiles at 880 and 1330\,$\mu$m.  Error bars represent the 
first and third quartiles at each wavelength or spatial frequency distance.  
The visibility profiles are normalized at a spatial frequency distance of 
$\sim$20\,k$\lambda$, and no correction is attempted to account for the 
slightly different distances to the two star-forming regions where the sample 
disks are located.  Overlaid on the median datapoints are the accretion disk 
models described above.  An examination of the evolutionary behaviors of 
submillimeter emission, mass accretion rates, and disk sizes in Figure 
\ref{evolve} and the corresponding typical SED and spatial distribution of 
emission in Figure \ref{median_data} indicate that the $\gamma = 1$ similarity 
solution for a fiducial viscous accretion disk model with $\alpha = 0.01$ 
generally describes a wide range of observed disk properties remarkably well.  
The fiducial model with a lower value of $\alpha$ (i.e., less turbulent 
viscosity) tends to over-predict submillimeter fluxes and accretion rates and 
under-predict disk sizes.  While an increase of $R_1$ to $\sim$100\,AU in this 
case would help in explaining the observed disk sizes and spatial emission 
distribution, it would also result in very little evolution in both the 
submillimeter fluxes and accretion rates, contrary to the observed trends.  The 
reader is referred to the work of \citet{hartmann98} regarding the effects of 
variations in other parameters.  We conclude that viscous accretion disks with 
$\alpha = 0.01$ provide the best general description of the observational 
properties of the disks in this sample.  

\subsection{Planet Formation}
In terms of utilizing the density structure of circumstellar disks to better 
understand the planet formation process, we can compare with the MMSN and a 
similar minimum mass ``extrasolar" nebula \citep[MMEN;][]{kuchner04} as 
references.  Unfortunately, these are not necessarily valuable comparisons.  
First, the interferometer data and MMSN/MMEN are providing information about 
different regions in disks; a direct analogy between the two is still limited 
by the spatial resolution of the data.  And second, while the circumstellar 
disk observations provide a measurement of $\Sigma_r$ at one instant in an 
evolution sequence, the MMSN and MMEN refer to the density structure 
\emph{integrated over the entire evolutionary history} of the solar and 
extrasolar disks.  In the later evolutionary stages when planetary embryos have 
formed and can dynamically influence the disk structure, it would not be so 
surprising if the density distribution varied significantly from the $p 
\approx 1$ expected for a steady-state accretion disk.  

In that sense, it is not too much of a concern that the inferred $p$ values for 
disks generally appear to be different than the $p \approx 1.5-2$ based on a 
fit to augmented planetary masses 
\citep{weidenschilling77,hayashi85,kuchner04}.  Moreover, it has been suggested 
that a double power-law behavior of the MMSN density distribution is more 
reasonable.  For example, \citet{lissauer87} points out another scenario where 
$p \approx 0.5$ for radii less than $\sim$10\,AU and $p \approx 1$ for larger 
radii.  This prescription is in rather good agreement with detailed accretion 
disk models \citep{dalessio98,dalessio99}.  \citet{raymond05} point out that, 
as long as there is sufficient material available, the \emph{ability to form 
planets} does not strongly depend on $p$.  Therefore, the real issue of 
comparison between the observations of circumstellar disks and the requirements 
of planet formation models lies more with the normalization, and not the shape, 
of the density distribution.  The important constraint is on how much mass is 
packed into the area where planets are thought to form.

Regardless of whether planets are created by core accretion and gas capture 
\citep[e.g.,][]{pollack96,inaba03}, gravitational instability 
\citep[e.g.,][]{boss98,boss05}, or a hybrid scenario 
\citep[e.g.,][]{durisen05}, all of these models require total masses in the 
inner disk at least one order of magnitude higher than the values inferred here 
from the constraints on $\Sigma_r$.  This is illustrated schematically in 
Figure \ref{planetformation}, where we have used the best-fit values of 
\{$\Sigma_5$, $p$\} in Table \ref{structure_table} to plot cumulative disk 
masses (the total mass internal to radius $r$) as a function of radius for the 
disks in this sample.  The shaded region in this figure corresponds to the 
surface density parameters required to provide at least $\sim$0.1\,M$_{\odot}$ 
of material within a radius of 20\,AU, the typical prerequisite values for 
current planet formation models.  Taking the disk structure constraints at face 
value would suggest that these disks will not be capable of forming planets on 
reasonable timescales.  However, this is not necessarily the case.  As we have 
emphasized above, the density distribution is currently only measured in the 
outer disk (beyond $\sim$60\,AU at best) and the densities in the inner disk 
are extrapolated from these measurements assuming a very simple structure 
model.  Without higher resolution data capable of tracing the inner disk 
material, we simply do not have a direct constraint on densities in the regions 
of interest for planet formation.  

There is one particularly notable way to reconcile the observationally inferred 
densities and those required by planet formation models: adjustments to the 
disk opacity.  The analysis in \S 4.2 and a number of other studies indicates 
that dust grains in the outer disk have grown to millimeter-scale sizes.  The 
presence of a substantial population of \emph{centimeter}-scale solids has even 
been inferred for the outer regions of the TW Hya disk \citep{wilner05}.  
Considering the expected strong radial dependence of collision timescales in 
the disk, and thus the growth timescales modulo some coagulation probability, a 
significant population of even larger solids could exist near the midplane in 
the inner parts of these disks.  Obviously, without a proper accounting of 
larger solids and their distribution with radius in the disk, the opacity 
formalism adopted in our models is unfortunately primitive and insufficient.  
Because the growth process would serve to decrease the opacity, reproducing the 
observed levels of disk emission would require a substantial increase in the 
densities (particularly for the inner disk).  Indeed, the models by 
\citet[][their Fig.~3]{dalessio01} indicate significantly decreased opacities 
when the maximum grain size is larger or smaller than a few millimeters.  Of 
equal importance is the implicitly assumed mass ratio of gas to dust in the 
disk midplane.  Here we assume the dust phase contributes only $\sim$1\%\ of 
the mass, as in the interstellar medium.  However, given the preferential 
sedimentation of larger solids to the midplane, it would not be surprising if 
the solid mass fraction in the regions responsible for the continuum emission 
was substantially higher.  Larger solid mass fractions would also imply 
proportionately smaller opacities, and thus higher densities \citep[see 
also][]{youdin02}.

In light of all this, it is likely that our extrapolated estimates of densities 
in the inner disk are significantly underestimated, at least in part due to a 
vastly over-simplified prescription for the opacity.  If the average opacity is 
roughly an order of magnitude lower than we assume, the densities in the 
typical disk should be sufficient to form planetary systems (the issue then 
becomes deciding which formation mechanism is more likely).  Unfortunately, 
diagnosing the effects of grain growth on the observable data for any 
individual disk is a daunting task.  One has to be concerned not only with the 
complicated growth and sedimentation processes 
\citep{dalessio01,dalessio06,dullemond04,dullemond05}, but also the structural, 
mineralogical, and dynamical properties of the aggregate solids 
\citep[e.g.,][]{wright87,henning95,takeuchi05}.  So, while we might not be able 
to determine with great certainty the magnitude of the effects of these 
processes on the inferred disk densities, the sense that they are 
underestimated is fairly certain.

\subsection{Future Work}
The prospects for significant improvements on basic disk structure measurements
are excellent.  The current suite of interferometers (SMA, PdBI, NMA, VLA) 
operate in all of the major atmospheric windows beyond $\sim$450\,$\mu$m with a 
typical spatial resolution of $\sim$1\arcsec\ and the potential to improve up 
to a factor of $\sim$3 higher.  The CARMA interferometer will provide a 
substantial increase in sensitivity and resolution, where its kilometer-scale 
baseline lengths will allow direct probes of the density into the inner regions 
($r\lesssim20$\,AU) of nearby disks.  We have briefly explored the potential 
advances in constraining the power-law index of the density distribution ($p$) 
via modeling simulated flat disk data for various observational facilities.  
For a typical disk with $p = 1$ and $R_d = 200$\,AU, the standard setups for 
current interferometers can potentially measure $p$ to within roughly 
$\pm$0.3$-$0.4 (e.g., see the WaOph 6 disk in Figure \ref{data_models6} with 
the SMA C2+E configuration).  The accuracy of the constraint on $p$ basically 
increases linearly with the spatial resolution, indicating that kilometer-scale 
CARMA baselines should be capable of measuring $p$ to within $\pm$0.1.  The 
added benefit of improved spatial resolution is the sensitivity to the density 
distribution on ever smaller spatial scales in the disk, probing into the 
region where giant planets are expected to form.  The real key to realizing 
this potential lies primarily with the ability to make phase corrections on the 
longest baseline separations (e.g., using water-vapor radiometry).  

The promise of submillimeter data with higher sensitivity and resolution will 
also allow some of the restrictions in the modeling described here to be 
relaxed.  The obvious goal will be to switch some of the fixed parameters into 
free ones (for example, fitting for the disk geometry directly from the 2-D 
distribution of visibilities).  Of particular interest in this case is letting
the power-law index of the opacity spectrum ($\beta$) be a free parameter, 
rather than relying on the approximate correction to the observed SED slope 
discussed in \S 4.2.  Ideally, this would be done in a direct way by adopting 
a model adjustment to simultaneously fit interferometric data at multiple 
wavelengths \citep[e.g.,][]{lay97}.  Almost all interferometers have a 
simultaneous dual-wavelength capability well-suited for this kind of analysis.  

\section{Summary}
We have used the SMA interferometer to conduct a high spatial resolution 
submillimeter continuum survey of 24 circumstellar disks in the Taurus-Auriga 
and Ophiuchus-Scorpius star formation regions.  By simultaneously fitting the 
broadband SEDs and submillimeter continuum visibilities to a simple disk model, 
we have placed constraints on the basic structural properties of these disks.  
We find:
\begin{enumerate}
\item The radial distributions of temperature and density in these disks have 
been determined, assuming they follow single power-law behaviors.  The median 
temperature and surface density profiles for the sample behave as $T_r \approx 
200\,r^{-0.62}$\,K and $\Sigma_r \approx 31\,r^{-0.5}$\,g cm$^{-2}$, 
respectively, where $r$ is measured in AU.  However, possible systematic errors 
and a more realistic temperature distribution in the outer disk indicate that 
the surface density probably has a steeper drop-off than is directly inferred 
here, with a typical power law index $p \approx 0.7-1.0$.  The distribution of 
the outer radii measured for these disks peaks at $R_d \approx 200$\,AU.

\item The inferred density distributions for these disks are consistent with 
similarity solutions for steady accretion disks, where the viscosity has a 
linear dependence on the radius and the $\alpha$ parameter is roughly constant 
\citep[see][]{hartmann98}.  A fiducial viscous evolution model with $\alpha 
\approx 0.01$ provides an excellent match to the variations of submillimeter 
flux densities, mass accretion rates, and outer radii with stellar age, as well 
as the typical sample SED and submillimeter surface brightness distribution.

\item Using the disk structure measurements and the shape of the submillimeter 
SED, we derive constraints on the power-law index of the opacity spectrum, 
$\beta$.  Incorporating the appropriate corrections for optically thick 
emission, the median value for this sample is $\beta \approx 0.7$.  Theoretical 
models suggest that such low values of $\beta$ indicate that solids in the disk 
have grown to roughly millimeter sizes.  

\item Direct constraints on the likelihood of planet formation in these disks 
must await higher resolution observations.  The density distributions inferred 
for the disks in this sample are significantly more shallow than those 
calculated for the MMSN/MMEN and used in many planet formation models.  
Extrapolating these distributions into the inner disk indicates that either too 
little material is available to form planets by the traditional mechanisms or, 
more likely, the adopted opacity prescription leads to significant density 
(i.e., mass) underestimates.  
\end{enumerate}

\appendix
\section{Comments on Individual Sources}
The literature sources of the continuum flux densities used to compile SEDs for 
this sample are given in Table \ref{seds_table}.  Some brief additional 
commentary for individual objects follows:

\emph{04158+2805} --- This object has the optical/infrared scattered light 
pattern consistent with a high inclination disk, and is also seen as a large
silhouette in the foreground of some nebulosity (F. M\'{e}nard, private 
communication).  Our modeling indicates a very large disk with a low value of 
$p$, despite the facts that the inclination is well-known and the best-fit 
temperature profile is similar to that expected for a flared disk.  The 
continuum image shows a peak offset from the phase center with an integrated 
flux significantly lower than the single-dish value.  Given the best-fit model, 
we find that these discrepancies are consequences of both the noise and the 
spatial filtering of such a large disk with a shallow surface brightness 
profile.  This is truly the most anomalous individual disk in the sample, 
especially considering the presumed low mass of the central star 
($\sim$0.1\,M$_{\odot}$ or less).  

\emph{AA Tau} --- Extensive photometric, spectroscopic, and polarimetric 
monitoring of this source indicate that its disk is also viewed at a high 
inclination, and may even require a warp to explain the time variability of the 
data \citep[e.g.,][]{bouvier99,osullivan05}.  The simple approximation of 
radiative transfer used in the flat disk model for the most inclined objects 
should be treated with some caution \citep[see][]{chiang99}.  

\emph{DH Tau} --- The DH Tau (A) primary has a nearby brown dwarf companion 
\citep{itoh05} which is not detected here.  With a projected separation of 
$\sim$320\,AU, the calculations by \citet{artymowicz94} suggest that the DH Tau 
(A) disk might be truncated at a radius of $\sim$0.2$-$0.4 times that 
separation, or $R_d \approx 60-120$\,AU.  This is in fairly good agreement with 
the fact that the DH Tau (A) disk is unresolved in these observations, although 
the noise in the visibilities is high enough that we can not statistically rule 
out a larger radius.  No disk around the nearby star DI Tau was detected in our 
observations, with a 3\,$\sigma$ upper limit of $\sim$10\,mJy at 880\,$\mu$m.  

\emph{DM Tau} --- \citet{calvet05} conclude that this disk has an inner hole 
out to $\sim$3\,AU in radius.  This requires us to modify the modeling 
procedure, as described in \S 3.3.  The flat disk model alone is unable to 
reproduce the mid-infrared emission in the SED, which is explained by 
\citet{calvet05} as due to a puffed-up inner disk rim which is directly 
irradiated by the central star.  Our modeling concentrates on the outer disk 
only, and therefore the results should be considered with caution (see also GM 
Aur and DoAr 25).  

\emph{DN Tau} --- The inclination value used in the modeling here ($i = 
28$\degr), determined based on magnetospheric accretion models 
\citep{muzerolle03}, is significantly lower than the value based on an 
elliptical Gaussian fit to the visibilities ($i \approx 65$\degr: see Table 
\ref{cont_table}).  However, running the modeling procedure with the latter 
inclination does not lead to significantly different parameter values.  

\begin{figure}
\epsscale{1.0}
\plotone{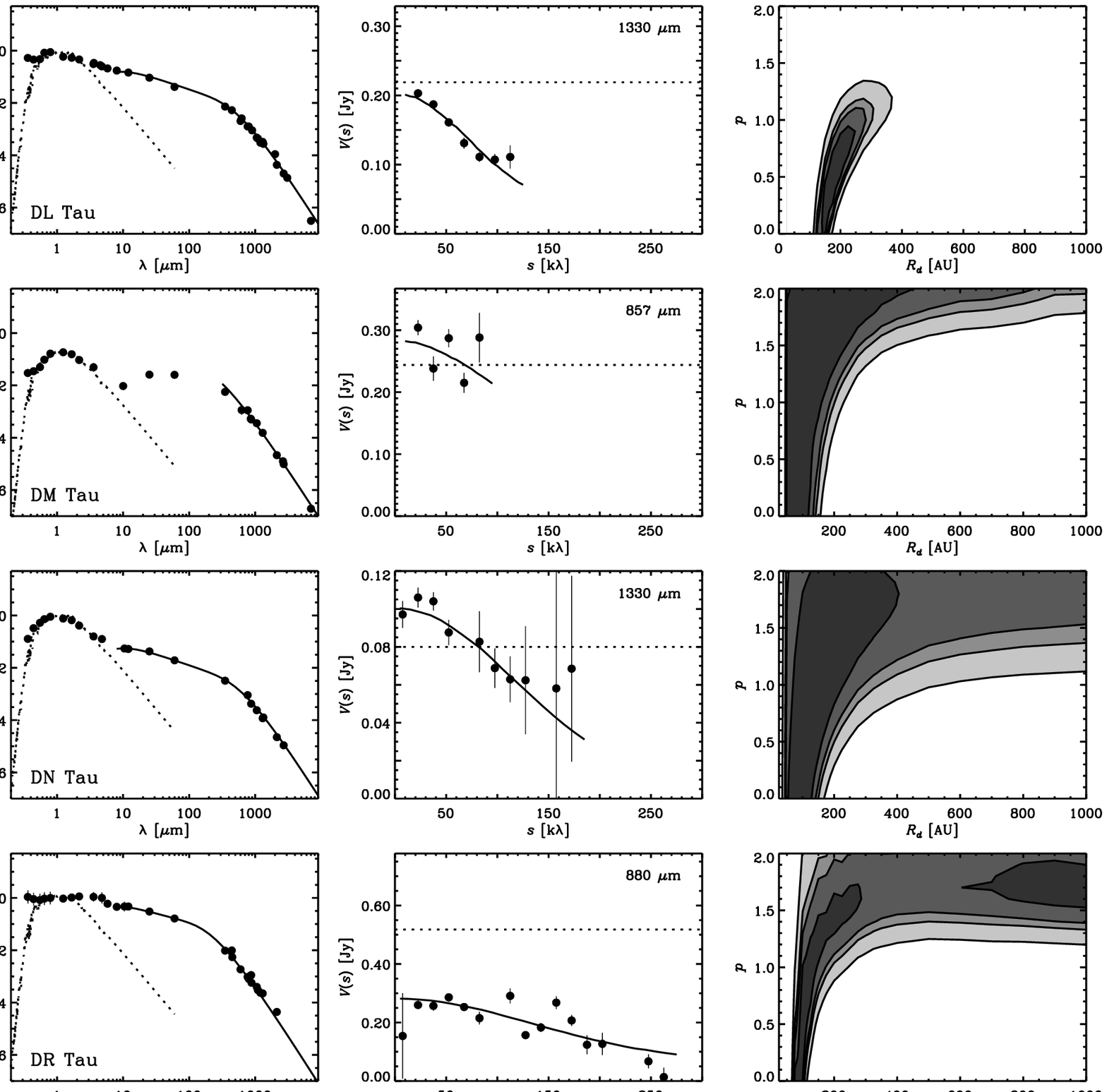}
\vspace{0.2cm}
\figcaption{Same as Fig.~\ref{data_models1}.\label{data_models2}}
\end{figure}

\begin{figure}
\epsscale{1.0}
\plotone{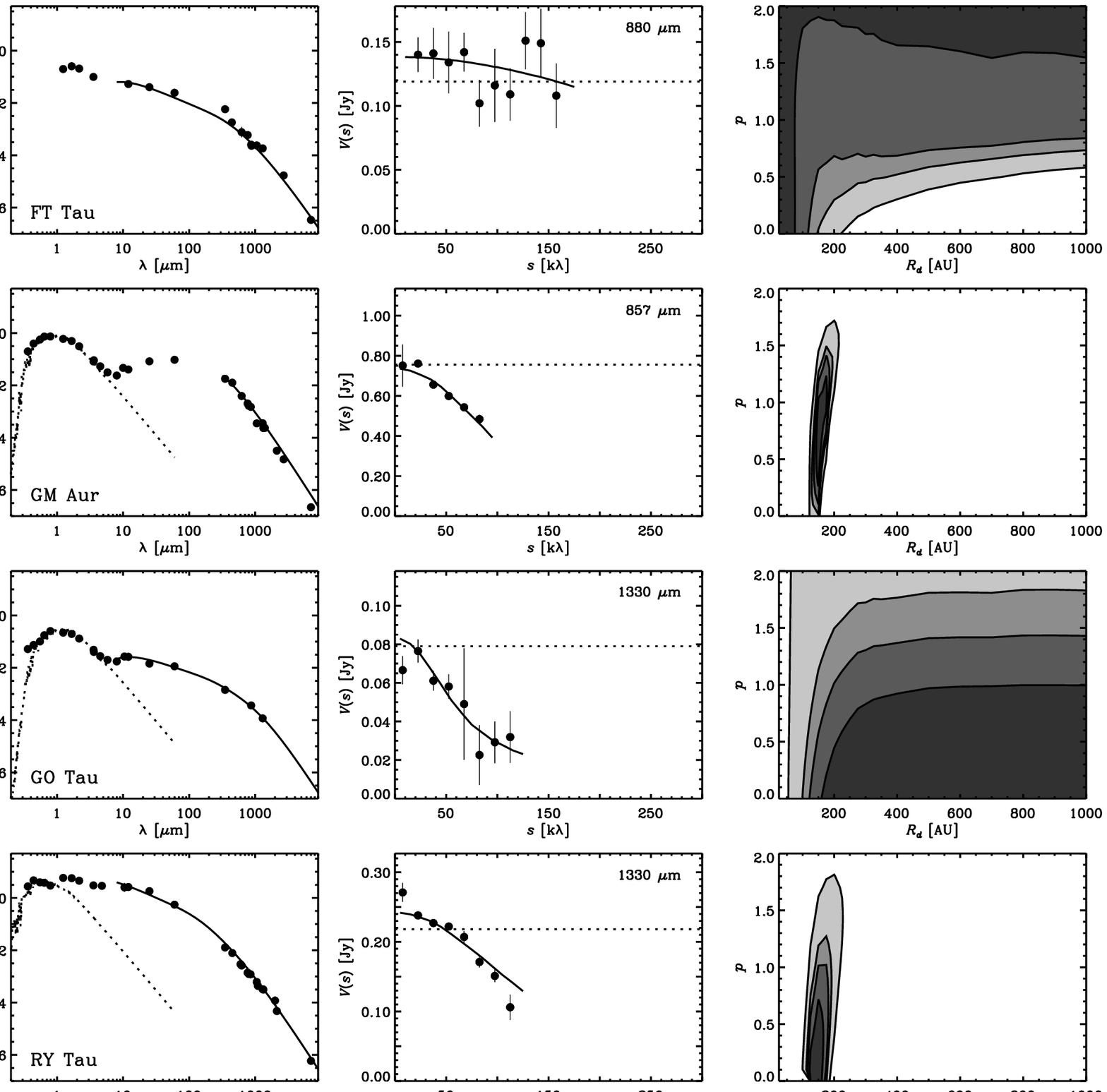}
\vspace{0.2cm}
\figcaption{Same as Fig.~\ref{data_models1}.\label{data_models3}}
\end{figure}

\begin{figure}
\epsscale{1.0}
\plotone{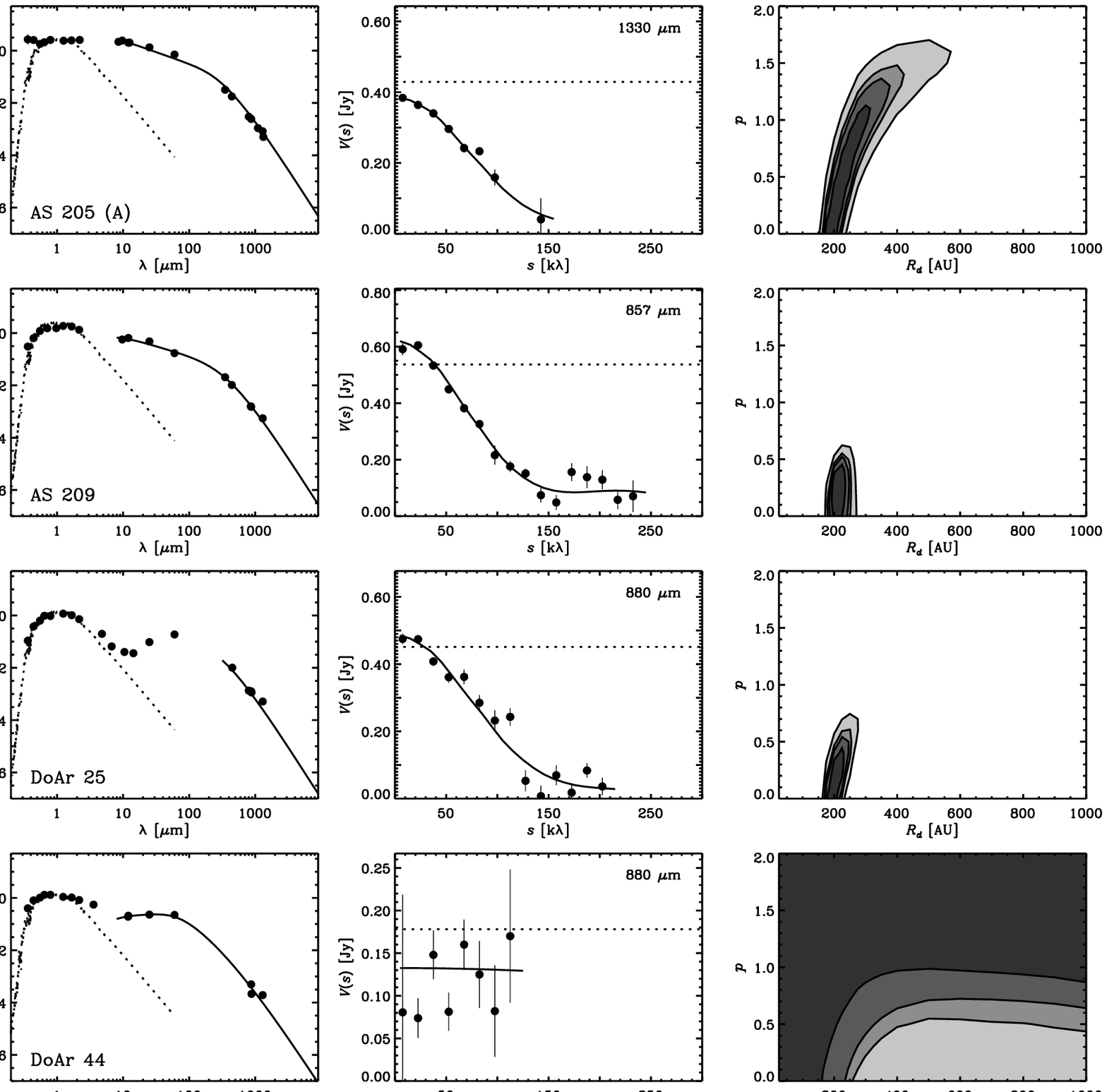}
\vspace{0.2cm}
\figcaption{Same as Fig.~\ref{data_models1}.\label{data_models4}}
\end{figure}

\begin{figure}
\epsscale{1.0}
\plotone{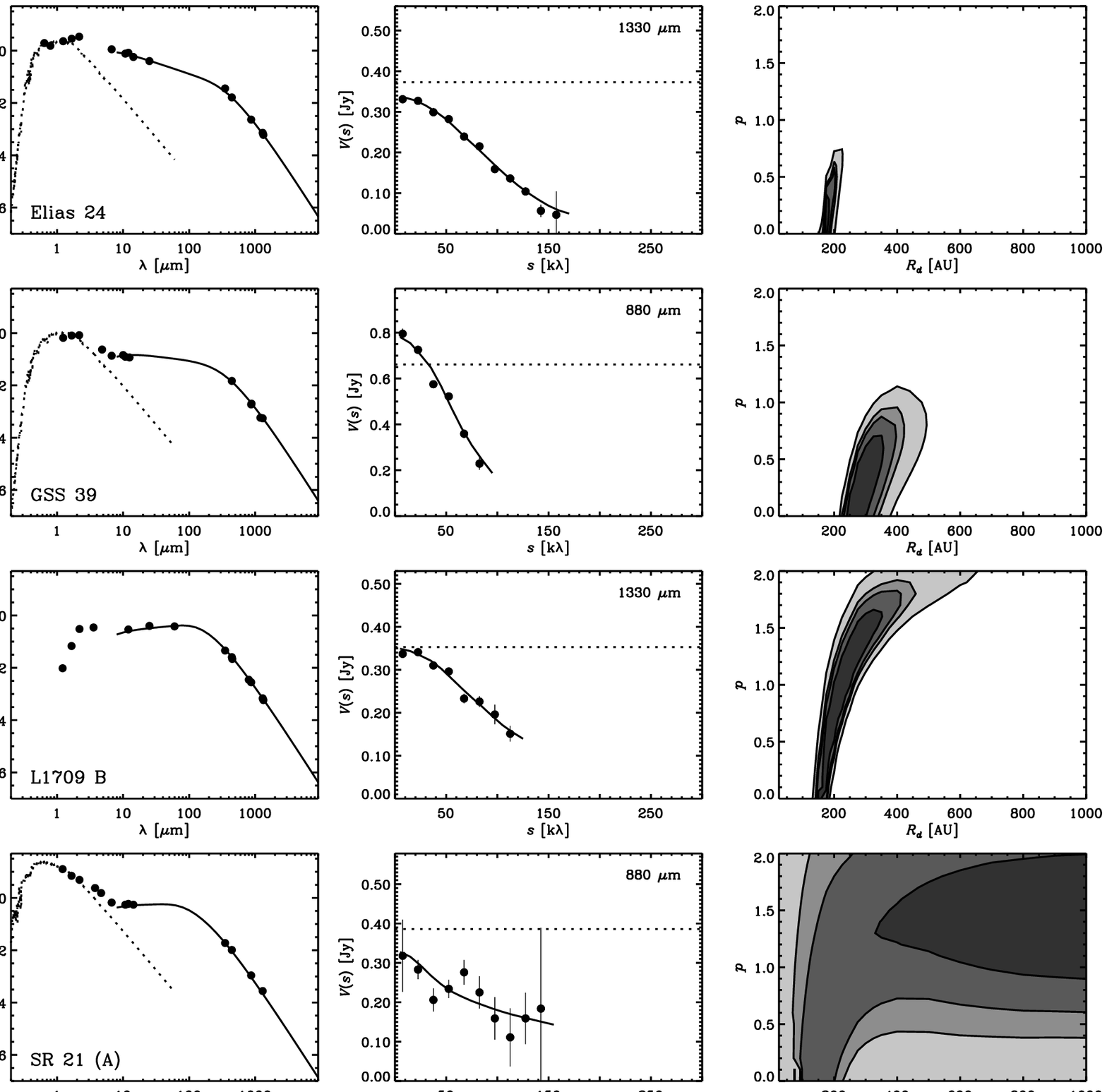}
\vspace{0.2cm}
\figcaption{Same as Fig.~\ref{data_models1}.\label{data_models5}}
\end{figure}

\begin{figure}
\epsscale{1.0}
\plotone{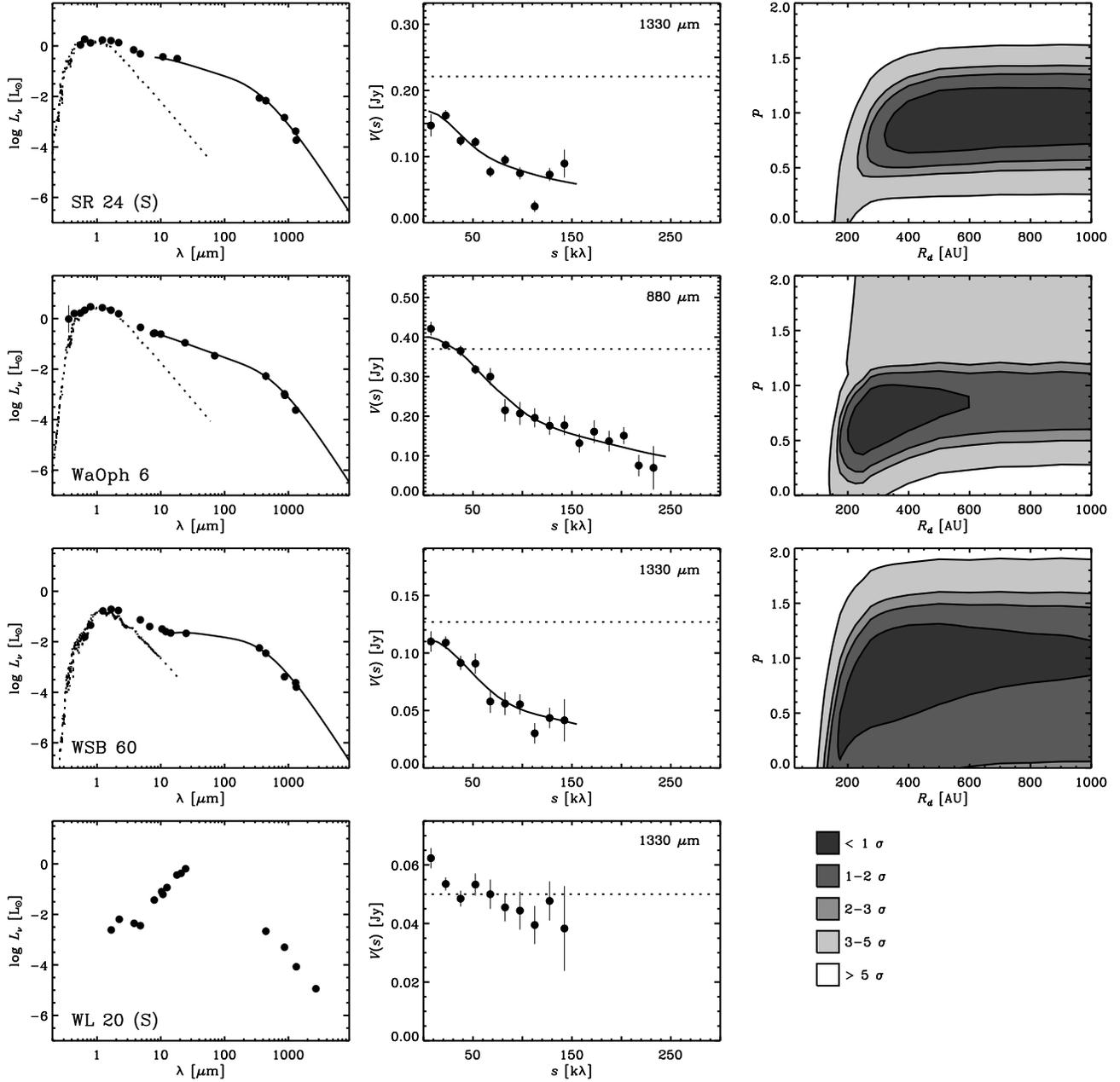}
\vspace{0.2cm}
\figcaption{Same as Fig.~\ref{data_models1}.\label{data_models6}.  Since no fit 
is performed for the Class I object WL 20 (S), we again show a key to the 
grayscale/contours in place of the $\Delta \chi^2$ map.}
\end{figure}

\begin{figure}
\epsscale{0.5}
\plotone{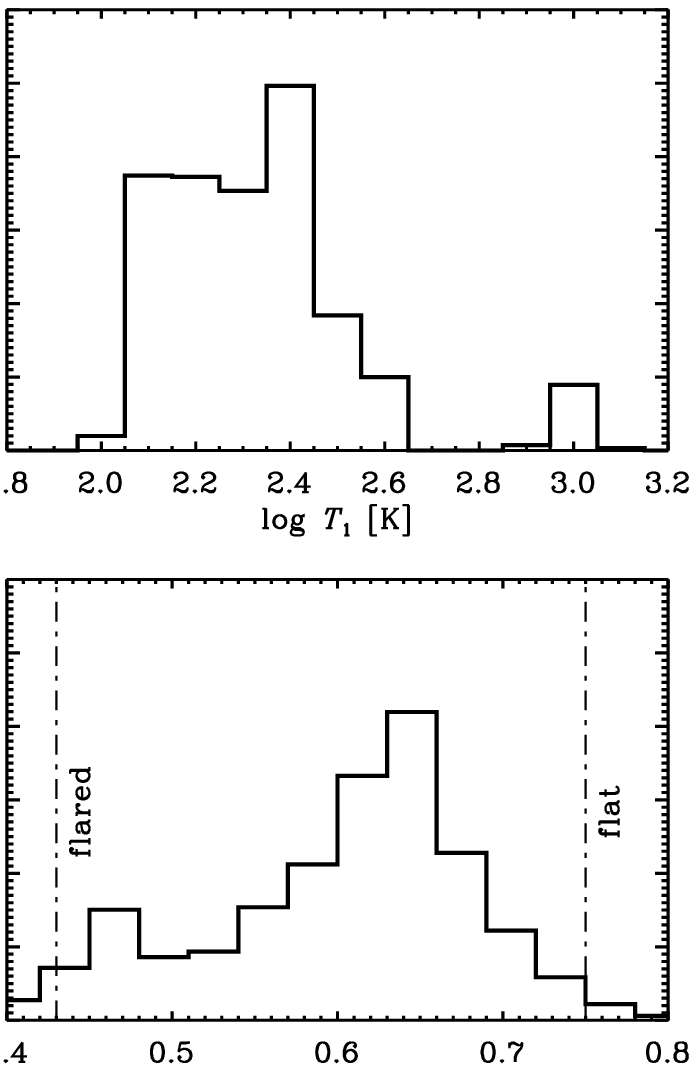}
\vspace{0.5cm}
\figcaption{Distributions of the parameters describing the radial variation of 
disk temperatures for this sample: the top panel for $\log{T_1}$, the 
temperature at 1\,AU, and the bottom panel for $q$, the radial power-law 
index.  The distributions were created to roughly account for parameter 
uncertainties as described in \S 4.1.1 for each measurement in Table 
\ref{structure_table}.  The bottom panel labels the idealized $q$ values for 
flared ($q\approx0.43$) and flat ($q\approx0.75$) disks.  \label{Tr_dists}}
\end{figure}

\begin{figure}
\epsscale{0.8}
\plotone{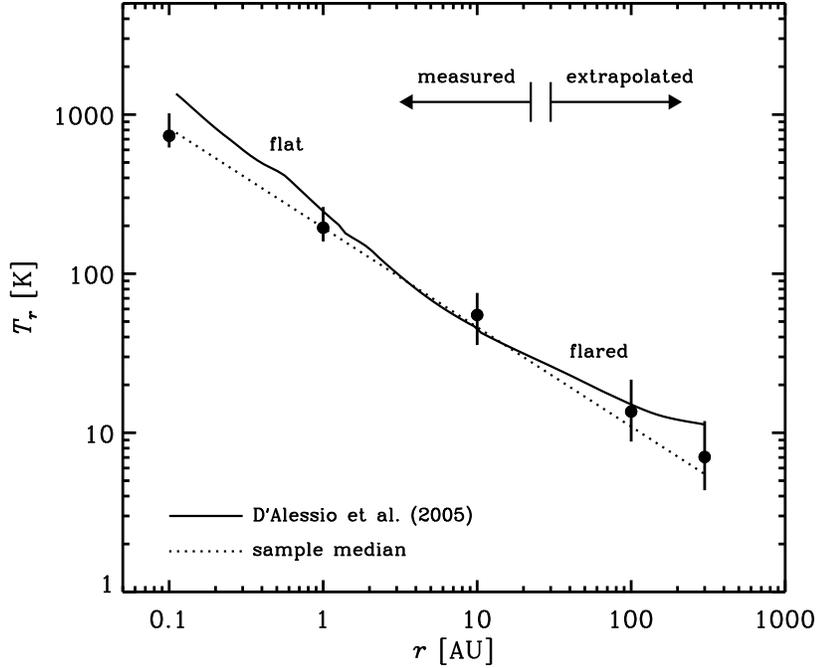}
\figcaption{A median temperature distribution for the sample.  The datapoints 
correspond to sample median temperatures at representative radii determined 
from the best-fit \{$T_1$, $q$\} values in Table \ref{structure_table}.  Error 
bars show the first and third quartiles of the temperatures at each radius, 
meant to represent the range of $T_r$ in the sample.  The dotted line gives the 
temperature profile constructed from the median values of \{$T_1$, $q$\}.  The 
solid curve shows the distribution of the midplane disk temperature from a more 
detailed model \citep[cf.,][]{dalessio05}, with star/disk parameters 
representative of the sample (see text).  Such models exhibit temperature 
distributions which are approximately flat for small radii and flared for 
larger radii, as labeled here.  A reminder at the top of the plot distinguishes 
the regions in the disk where the temperature distribution is actually measured 
(inner disk) or merely extrapolated (outer disk). \label{Tr}}
\end{figure}

\begin{figure}
\epsscale{0.5}
\plotone{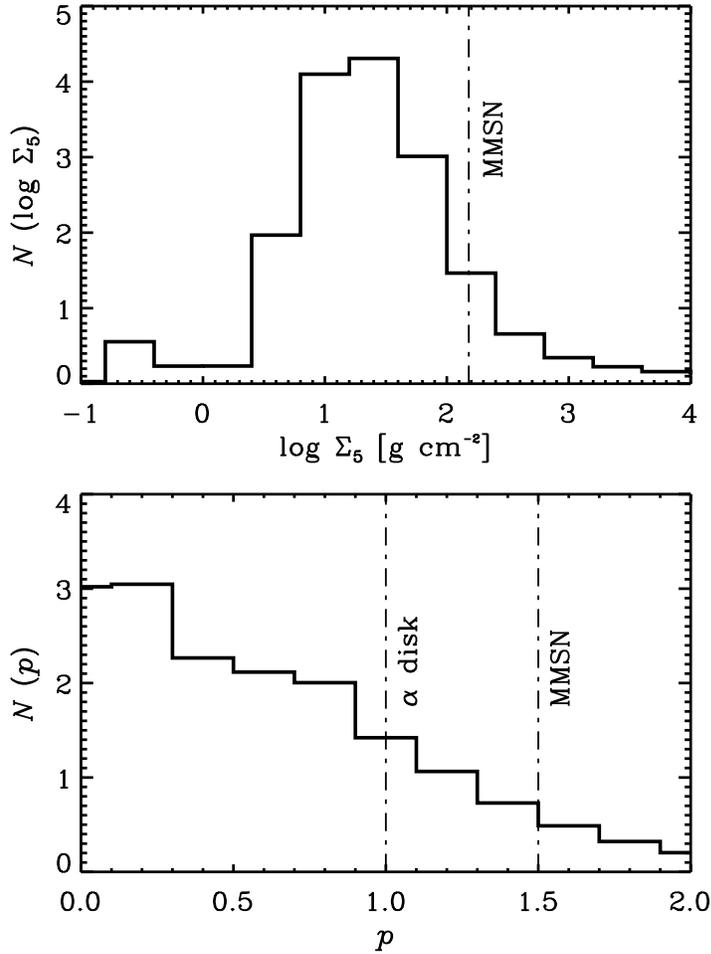}
\vspace{0.5cm}
\figcaption{Distributions of the disk surface density profile parameters for 
this sample: the top panel for $\log{\Sigma_5}$, the surface density at 5\,AU, 
and the bottom panel for $p$, the radial power-law index.  The distributions 
were created as for Fig.~\ref{Tr_dists} based on the best-fit values in Table 
\ref{structure_table}.  The top panel shows the 5\,AU surface density for the 
minimum mass solar nebula (MMSN), and the bottom panel shows the power-law 
indices for both the MMSN and a steady viscous accretion disk.\label{SDr_dists}}
\end{figure}

\begin{figure}
\epsscale{0.8}
\plotone{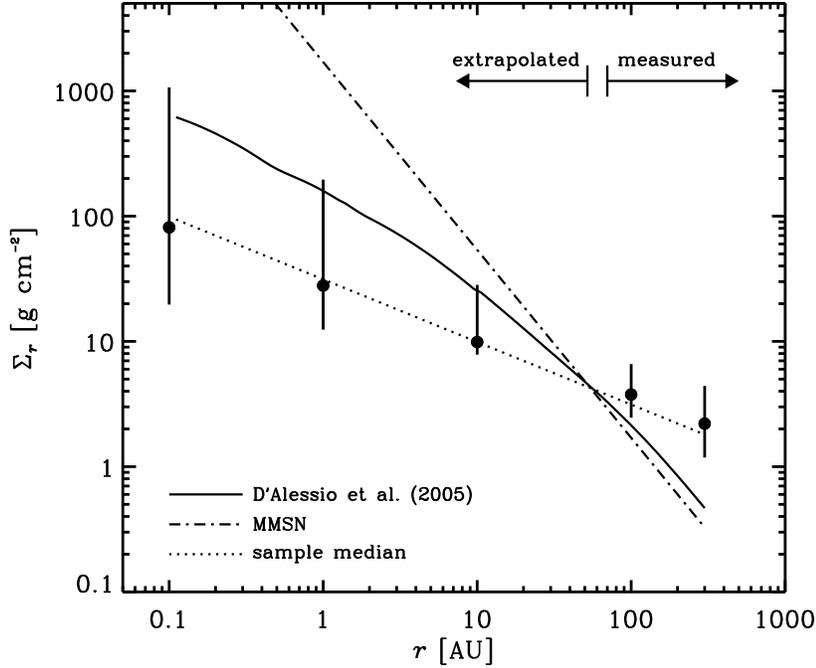}
\figcaption{A median surface density distribution for the sample.  As in 
Fig.~\ref{Tr}, the datapoints correspond to sample median surface densities at 
representative radii determined from the best-fit \{$\Sigma_5$, $p$\} values in 
Table \ref{structure_table}.  Error bars show the first and third quartiles of 
the surface densities at each radius, meant to represent the range of 
$\Sigma_r$ in the sample.  The dotted line gives the surface density profile 
constructed from the median values of \{$\Sigma_5$, $p$\}.  The solid curve 
shows the density distribution from a detailed model 
\citep[cf.,][]{dalessio05}, with the same parameters as in Fig.~\ref{Tr} (see 
text).  Such models have $\Sigma_r \propto r^{-1}$ and an exponential cut-off 
at large radii.  The dot-dashed line marks the MMSN density distribution.  A 
reminder at the top of the plot distinguishes the regions in the disk where 
$\Sigma_r$ is actually measured (outer disk) or merely extrapolated (inner 
disk).  \label{SDr}}
\end{figure}

\begin{figure}
\epsscale{0.5}
\plotone{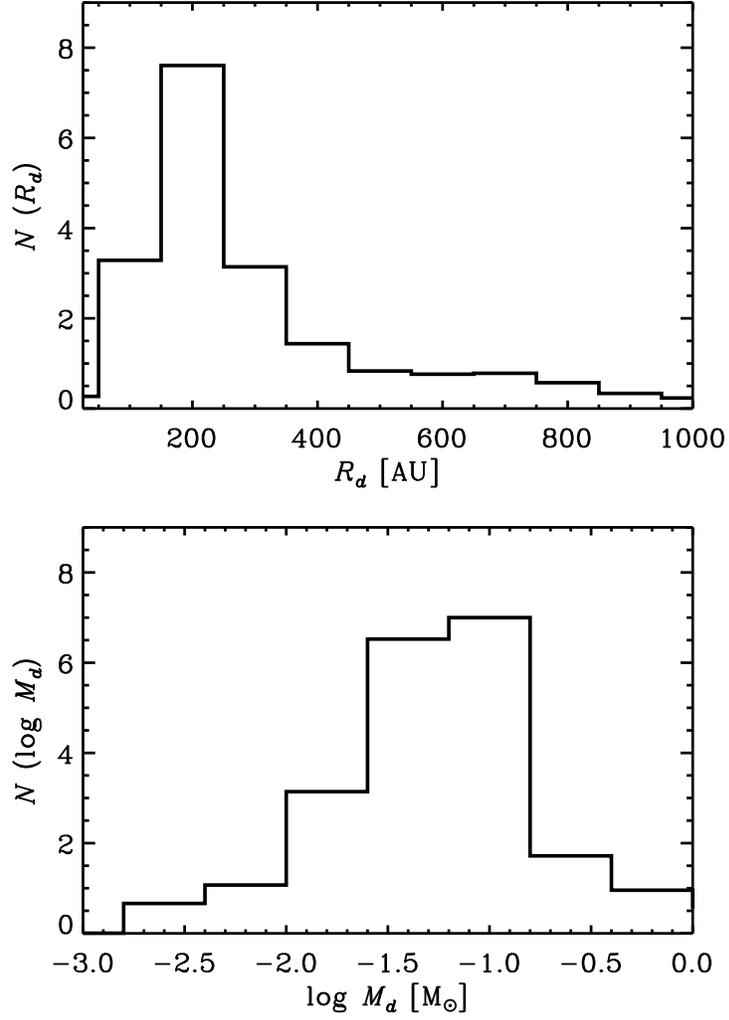}
\vspace{0.5cm}
\figcaption{Distributions of the outer disk radii, $R_d$, and total disk 
masses, $M_d$, for this sample, created as for those in Figs.~\ref{Tr_dists} 
and \ref{SDr_dists} from the measurements in Table \ref{structure_table}.  
\label{RM_dist}}
\end{figure}

\begin{figure}
\epsscale{0.5}
\plotone{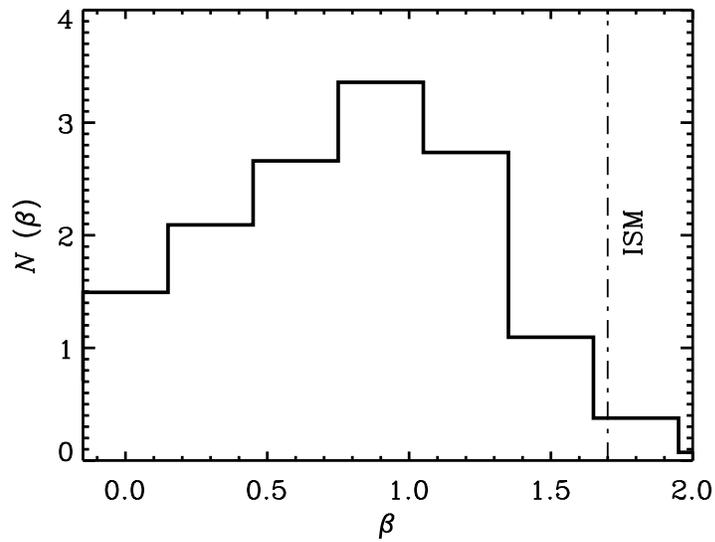}
\vspace{0.5cm}
\figcaption{Distribution of the power-law index of the opacity spectrum, 
$\beta$, created as for Figs.~\ref{Tr_dists}, \ref{SDr_dists}, and 
\ref{RM_dist} from the values listed in Table \ref{opacity_table}.  The value 
of $\beta$ for the ISM is marked with a dot-dashed vertical line. 
\label{beta_dist}}
\end{figure}

\begin{figure}
\epsscale{1.0}
\plotone{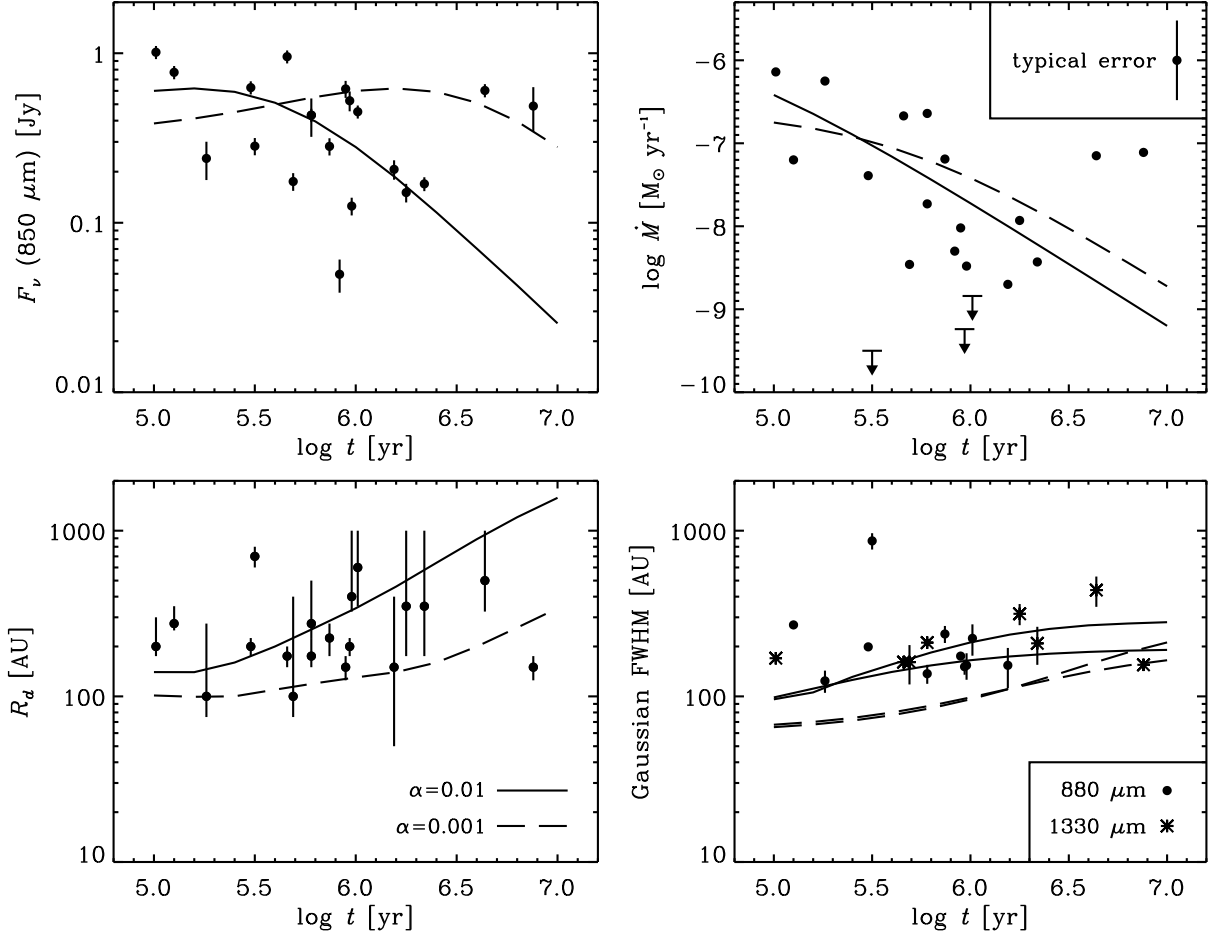}
\vspace{0.3cm}
\figcaption{Variations of the 850\,$\mu$m flux densities (\emph{top left}), 
mass accretion rates (\emph{top right}; from the literature, see Table 
\ref{stars_table}), outer radii (\emph{bottom left}), and elliptical Gaussian 
FWHM (\emph{bottom right}; determined from a fit to the C1 or C2 visibilities) 
with stellar age (see Table \ref{stars_table}).  Overlaid are the expected 
trends for two fiducial accretion disk models with $\alpha = 0.01$ (solid) and 
$\alpha = 0.001$ (dashed).  Descriptions of the other parameters and the method 
of relating flat disk $R_d$ values with the accretion disk models are given in 
\S 5.1.  The different symbols in the bottom right panel describing the FWHM 
correspond to two observing wavelengths used in this survey.  The double 
accretion disk model curves also correspond to these two wavelengths, where the 
1330\,$\mu$m FWHM values are always slightly larger than at 880\,$\mu$m.  
Typical systematic errors on stellar ages could be a factor of 2 or more for 
any individual object.  \label{evolve}}
\end{figure}

\begin{figure}
\epsscale{0.4}
\plotone{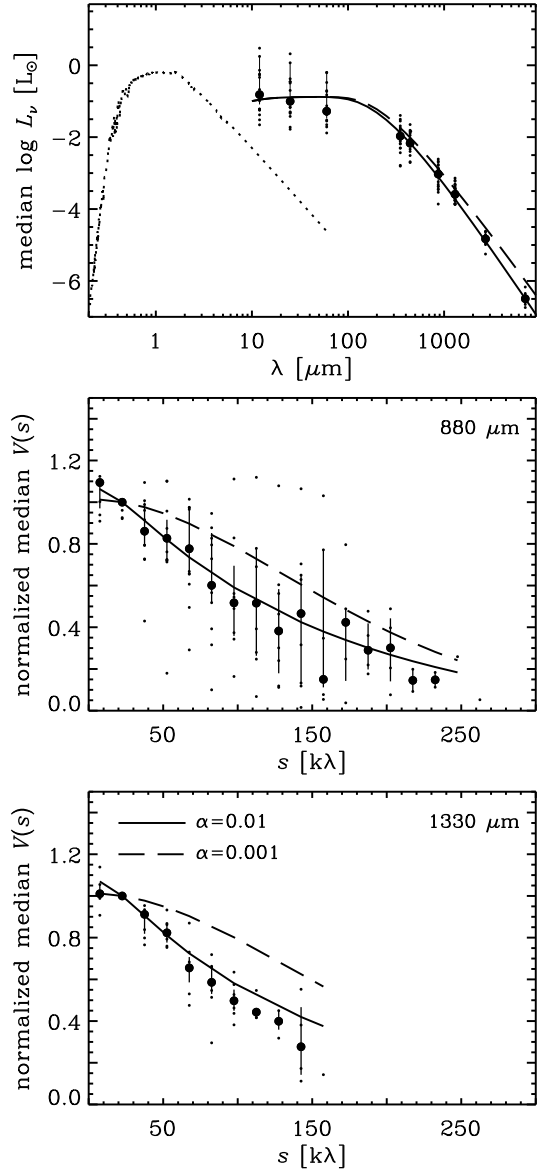}
\vspace{0.3cm}
\figcaption{The sample median SED (\emph{top}) and flux-normalized visibility 
profiles at 880\,$\mu$m (\emph{middle}) and 1330\,$\mu$m (\emph{bottom}) are 
shown to approximate the typical observables for a T Tauri disk.  The error 
bars indicate the first and third quartiles at each wavelength or spatial 
frequency distance.  The visibility profiles have their fluxes normalized at a 
spatial frequency distance $s \sim 20$\,k$\lambda$.  The data from individual 
disks are shown as small points.  The contribution of a K7 stellar photosphere 
is shown as a dashed profile in the top SED panel.  Overlaid on each panel are 
the same two fiducial accretion disk models shown in Figure \ref{evolve} for 
$\alpha = 0.01$ (solid) and $\alpha = 0.001$ (dashed).  These models were 
computed at the median age for the sample, $\sim$1\,Myr.  \label{median_data}}
\end{figure}

\begin{figure}
\epsscale{0.8}
\plotone{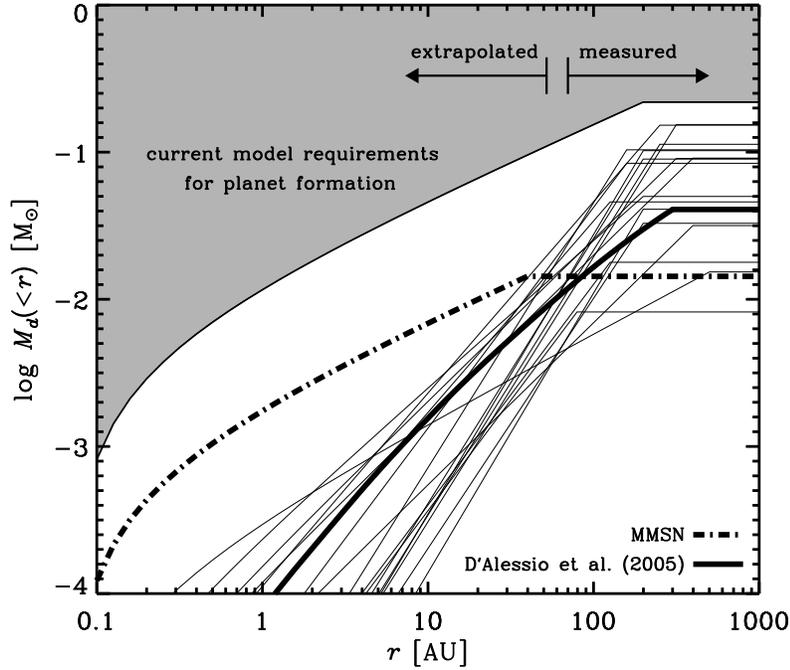}
\figcaption{Cumulative disk masses (i.e., the mass internal to $r$) as a 
function of radius for the sample (thin solid curves), determined with the 
best-fit surface density parameters listed in Table \ref{structure_table}.  The 
gray area roughly marks the density requirements for current planet formation 
models.  The lower boundary of this region corresponds to $\Sigma_5 = 10^3$\,g 
cm$^{-2}$ and $p = 1.5$ for a disk truncated at 200\,AU.  The heavy dashed 
curve marks the \citet{hayashi85} MMSN values with an outer radius of 50\,AU.  
The integrated surface density profile for the irradiated viscous accretion 
disk model calculated by \citet{dalessio05} and described in the text (see 
Fig.~\ref{SDr}) is also shown as a heavy solid curve.  The current 
observational estimates of densities in the disk fall short of the values 
required by planet formation models, perhaps in part due to an overestimate of 
opacities. \label{planetformation}}
\end{figure}

\emph{DR Tau} --- The visibility profile of the DR Tau disk (see 
Fig.~\ref{data_models2}) clearly shows that the single-dish observations 
include substantially more emission than the SMA data.  This may be due to 
extended emission that is resolved out with the interferometer, although 
additional observations would be required to make any definitive conclusions 
about the origin of that emission.  

\emph{GM Aur} --- As with the DM Tau disk, \citet{calvet05} suggest that GM Aur 
has a large inner hole out to $\sim$24\,AU in radius.  As described in \S 3.3, 
the usual fitting procedure was modified to fit only the outer disk.  The 
best-fit model for this disk is by far the poorest in the sample, predicting 
systematically higher submillimeter flux densities than are observed. 

\emph{RY Tau} --- The highest inclination disk in the sample.  The simple 
radiative transfer approximation used here is probably insufficient.  An 
alternative lower inclination ($i \approx 25$\degr) estimated from a CO 
spectrum \citep{koerner95} is most likely incorrect, given the inclination 
value estimated from elliptical Gaussian fits to the data presented here ($i 
\gtrsim 80$\degr), rotationally broadened stellar line profiles 
\citep[e.g.,][]{calvet04}, and extinction-related variability \citep{herbst99} 
observed for this source.

\emph{AS 205} --- The continuum emission in this multiple star system peaks at 
the position of the northern primary, AS 205 (A).  However, the orientation 
and extent of the best-fit disk model in this case would overlap in projection 
with the spectroscopic binary secondary system.  If the projected separation 
was similar to the true separation in this system, the AS 205 (A) disk should 
be dynamically truncated to less than half of the best-fit $R_d$ 
\citep[cf.,][]{artymowicz94}.  It is more likely that the AS 205 (B) system 
lies either in the foreground or background, and therefore has little or no 
influence on the observed submillimeter disk.

\emph{DoAr 25} --- The SED of this disk has substantially enhanced mid-infrared 
emission in the \emph{IRAS} bands, most likely due to contamination from the 
interstellar medium in the large beams.  Without the mid-infrared emission for 
this source, we again modifed the fitting procedure, as described in \S 3.3.  

\emph{SR 21} --- All of the disk emission described here is for the primary 
source, SR 21 (A).  Any disk around the binary companion has an 880\,$\mu$m 
flux density less than $\sim$20\,mJy (3\,$\sigma$).

\emph{SR 24} --- The disks in this triple system are discussed in detail in 
\citet{andrews05b}.  The continuum emission modeled here is for the isolated 
primary SR 24 (S).

\begin{deluxetable}{lccccccl}
\tablecolumns{8}
\tabletypesize{\scriptsize}
\tablewidth{0pc}
\tablecaption{Summary of SMA Observations\label{obssum_table}}
\tablehead{
\colhead{Object} & \colhead{$\alpha$ [J2000]} & \colhead{$\delta$ [J2000]} & \colhead{$\nu_{\rm{IF}}$} & \colhead{Config.} & \colhead{Beam Size} & \colhead{PA} & \colhead{UT Date(s)} \\ 
\colhead{} & \colhead{[$^{\rm{h}}$ $^{\rm{m}}$ $^{\rm{s}}$]} & \colhead{[\degr\, \arcmin\, \arcsec]} & \colhead{[GHz]} & \colhead{} & \colhead{[\arcsec]} & \colhead{[\degr]} & \colhead{} \\ 
\colhead{(1)} & \colhead{(2)} & \colhead{(3)} & \colhead{(4)} & \colhead{(5)} & \colhead{(6)} & \colhead{(7)} & \colhead{(8)}}
\startdata
04158+2805 & 04 18 58.1 & +28 12 23.5                    & 340.8 & C1     & $1.7\times1.3$ & 97  & 2004 Dec 13              \\
AA Tau     & 04 34 55.4 & +24 28 53.2                    & 340.8 & C1     & $1.8\times1.3$ & 95  & 2004 Dec 13              \\
CI Tau     & 04 33 52.0 & +22 50 30.2                    & 340.8 & C1     & $1.7\times1.0$ & 87  & 2004 Dec 10              \\
DH/DI Tau  & 04 29 42.0 & +26 32 53.2\tablenotemark{a}   & 340.8 & C2     & $2.3\times2.1$ & 51  & 2005 Dec 17              \\
DL Tau     & 04 33 39.1 & +25 20 38.2                    & 225.5 & C1     & $2.8\times1.7$ & 85  & 2004 Nov 27              \\
DM Tau     & 04 33 48.7 & +18 10 12.0\tablenotemark{b}   & 349.9 & C2     & $2.3\times2.1$ & 64  & 2005 Nov 26              \\
DN Tau     & 04 35 27.4 & +24 14 58.9                    & 225.5 & C2, E  & $2.1\times1.8$ & 89  & 2005 Nov 27, 2006 Jan 28 \\
DR Tau     & 04 47 06.2 & +16 58 42.9                    & 340.8 & E, C2  & $1.5\times1.2$ & 59  & 2005 Sep 9, Dec 17       \\
FT Tau     & 04 23 39.2 & +24 56 14.1                    & 340.8 & C1     & $1.7\times1.1$ & 94  & 2004 Dec 10              \\
GM Aur     & 04 55 10.9 & +30 22 01.0                    & 349.9 & C2     & $2.2\times2.1$ & 64  & 2005 Nov 26              \\
GO Tau     & 04 43 03.1 & +25 20 18.8                    & 225.5 & C2, E  & $2.1\times1.8$ & 87  & 2005 Nov 27, 2006 Jan 28 \\
RY Tau     & 04 21 57.4 & +28 26 35.5                    & 225.5 & C1     & $2.8\times1.7$ & 87  & 2004 Nov 27              \\
\hline
AS 205     & 16 11 31.3 & $-$18 38 25.9                  & 225.5 & E, C1  & $2.2\times1.7$ & 19  & 2004 Jun 15, Aug 9       \\
AS 209     & 16 49 15.3 & $-$14 22 08.7                  & 349.9 & C2, E  & $1.5\times1.2$ & 150 & 2006 May 12, Jun 3       \\
DoAr 25    & 16 26 23.6 & $-$24 43 13.2\tablenotemark{b} & 340.8 & E, C2  & $1.8\times1.6$ & 2   & 2005 May 8, Jun 12       \\
DoAr 44    & 16 31 33.5 & $-$24 27 37.3                  & 340.8 & C1     & $2.3\times1.4$ & 35  & 2004 Jul 31              \\
Elias 24   & 16 26 24.1 & $-$24 16 13.5                  & 225.5 & C2, E  & $2.2\times2.0$ & 159 & 2006 May 9, 27           \\
GSS 39     & 16 26 45.0 & $-$24 23 07.7                  & 340.8 & C2     & $3.4\times1.8$ & 98  & 2006 May 14              \\
L1709 B    & 16 31 35.7 & $-$24 01 29.5                  & 225.5 & E, C1  & $2.4\times1.8$ & 22  & 2004 Jun 15, Aug 9       \\
SR 21      & 16 27 10.3 & $-$24 18 12.7                  & 340.8 & C1     & $2.5\times1.4$ & 39  & 2004 Jul 31              \\
SR 24      & 16 26 58.5 & $-$24 45 36.9\tablenotemark{a} & 225.5 & C1, E  & $2.1\times1.3$ & 39  & 2004 Aug 2, 21           \\
WaOph 6    & 16 48 45.6 & $-$14 16 36.0                  & 340.8 & E, C2  & $1.6\times1.6$ & 30  & 2005 May 8, Jun 12       \\
WSB 60     & 16 28 16.5 & $-$24 36 58.0                  & 225.5 & E, C2  & $2.4\times1.9$ & 14  & 2005 May 15, Jun 28      \\
WL 20      & 16 27 15.9 & $-$24 38 43.4\tablenotemark{b} & 225.5 & C2, E  & $2.0\times1.8$ & 164 & 2006 May 9, 27           
\enddata
\tablecomments{Col.~(1): Object name.  Cols.~(2) \& (3): Phase center 
coordinates.  Col.~(4): IF observing frequency.  Col.~(5): SMA 
configuration(s); C1 = compact with some long baselines, C2 = compact, E = 
extended (see text).  Col.~(6): Dimensions of the naturally weighted 
synthesized beam made with all available configurations.  Col.~(7): Position 
angle of synthesized beam, measured east of north.  Col.~(8): UT date(s) of 
observation(s).} 
\tablenotetext{a}{Phase center coordinates set between the positions of objects 
in a multiple system.}
\tablenotetext{b}{Observed phase center coordinates are offset from emission 
peaks by a significant amount.  The position of DM Tau is 0\farcs25 to the 
east and 1\farcs9 to the south of the phase center.  The position of DoAr 25 
is 1\farcs8 to the east and 0\farcs8 to the south of the phase center.  All 
emission in the WL 20 system is from the southwestern component, located 
2\farcs6 to the west and 2\farcs3 to the south of the phase center.}
\end{deluxetable}

\clearpage

\begin{deluxetable}{lcrrr}
\tablecolumns{5}
\tabletypesize{\scriptsize}
\tablewidth{0pc}
\tablecaption{Continuum Measurements\label{cont_table}}
\tablehead{
\colhead{Object} & \colhead{$\lambda$} & \colhead{$F_{\nu}$} & \multicolumn{2}{c}{Gaussian fit parameters} \\
\colhead{} & \colhead{[$\mu$m]} & \colhead{[mJy]} & \colhead{FWHM size [\arcsec]} & \colhead{PA [\degr]} \\
\colhead{(1)} & \colhead{(2)} & \colhead{(3)} & \colhead{(4)} & \colhead{(5)}}
\startdata
04158+2805 & 880  & $67\pm2$  & $6.2\pm0.7\times3.7\pm0.7$     & $88\pm5$   \\
AA Tau     & 880  & $115\pm3$ & $1.1\pm0.2\times0.4\pm0.2$     & $94\pm9$   \\
CI Tau     & 880  & $216\pm3$ & $1.7\pm0.2\times1.1\pm0.1$     & $131\pm12$ \\
DH Tau (A) & 880  & $49\pm3$  & unresolved	                & \nodata   \\
DL Tau     & 1330 & $199\pm2$ & $1.51\pm0.08\times0.74\pm0.08$ & $44\pm5$   \\
DM Tau     & 857  & $249\pm3$ & $1.1\pm0.3\times0.18\pm0.08$   & $172\pm9$  \\
DN Tau     & 1330 & $90\pm2$  & $0.7\pm0.1\times0.3\pm0.2$     & $40\pm19$  \\
DR Tau     & 880  & $275\pm3$ & $0.61\pm0.05\times0.24\pm0.05$ & $170\pm8$  \\
FT Tau     & 880  & $111\pm2$ & unresolved                     & \nodata    \\
GM Aur     & 857  & $707\pm4$ & $1.25\pm0.05\times0.80\pm0.05$ & $58\pm4$   \\
GO Tau     & 1330 & $57\pm2$  & $2.0\pm0.3\times0.8\pm0.3$     & $0\pm10$   \\
RY Tau     & 1330 & $228\pm2$ & $1.11\pm0.07\times0.2\pm0.3$   & $32\pm5$   \\
\hline
AS 205 (A) & 1330 & $279\pm3$ & $1.00\pm0.03\times0.68\pm0.03$ & $55\pm7$   \\
AS 209     & 857  & $570\pm4$ & $1.14\pm0.03\times1.02\pm0.03$ & $35\pm13$  \\
DoAr 25    & 880  & $421\pm5$ & $1.28\pm0.06\times0.61\pm0.06$ & $112\pm3$  \\
DoAr 44    & 880  & $79\pm4$  & unresolved                     & \nodata    \\
Elias 24   & 1330 & $335\pm2$ & $1.09\pm0.02\times0.77\pm0.02$ & $27\pm3$   \\
GSS 39     & 880  & $736\pm5$ & $1.69\pm0.06\times1.10\pm0.08$ & $114\pm5$  \\
L1709 B    & 1330 & $325\pm3$ & $1.08\pm0.05\times0.62\pm0.05$ & $27\pm4$   \\
SR 21 (A)  & 880  & $227\pm7$ & $1.4\pm0.3$                    & $12\pm9$   \\
SR 24 (S)  & 1330 & $104\pm2$ & $1.4\pm0.1\times0.77\pm0.07$   & $25\pm5$   \\
WaOph 6    & 880  & $337\pm3$ & $0.77\pm0.04\times0.58\pm0.04$ & $12\pm8$   \\
WSB 60     & 1330 & $89\pm2$  & $1.2\pm0.1\times0.9\pm0.1$     & $3\pm14$   \\
WL 20 (S)  & 1330 & $47\pm1$  & $0.59\pm0.09\times0.30\pm0.09$ & $55\pm19$  
\enddata
\tablecomments{Col.~(1): Object name.  Col.~(2): Observing wavelength.  
Col.~(3): Integrated continuum flux density and 1\,$\sigma$ statistical error 
(systematic error due to absolute calibration uncertainty is not included; see 
text).  Col.~(4): FWHM dimensions of elliptical Gaussian fitted directly to the 
visibilties.  Col.~(5): Position angle (measured east of north) of elliptical 
Gaussian fit.}
\end{deluxetable}

\clearpage

\begin{deluxetable}{lcccccclc}
\tablecolumns{9}
\tabletypesize{\scriptsize}
\tablewidth{0pc}
\tablecaption{Sample Properties\label{stars_table}}
\tablehead{
\colhead{Object} & \colhead{SED} & \colhead{SpT} & \colhead{$A_V$} & \colhead{ref} & \colhead{log $t$} & \colhead{log \emph{\.{M}}} & \colhead{multiplicity} & \colhead{ref} \\
\colhead{} & \colhead{} & \colhead{} & \colhead{[mag]} & \colhead{} & \colhead{[yr]} & \colhead{[M$_{\odot}$ yr$^{-1}$]} & \colhead{} & \colhead{} \\
\colhead{(1)} & \colhead{(2)} & \colhead{(3)} & \colhead{(4)} & \colhead{(5)} & \colhead{(6)} & \colhead{(7)} & \colhead{(8)} & \colhead{(9)}}
\startdata
04158+2805 & II   & M6      & 8.6     & 1       & 5.50    & $\le -9.50$ & single (?)                  & \nodata \\
AA Tau     & II   & K7      & 1.1     & 2       & 5.98    & $-8.48$     & single                      & a \\
CI Tau     & II   & K7      & 1.8     & 2       & 5.87    & $-7.19$     & single                      & b \\
DH Tau (A) & II   & M1      & 1.3     & 2       & 5.92    & $-8.30$     & binary (2\farcs3)           & c \\
DL Tau     & II   & K7      & 2.1     & 2       & 5.78    & $-7.73$     & single                      & a \\
DM Tau     & II   & M1      & 0.6     & 2       & 6.19    & $-8.70$     & single                      & d \\
DN Tau     & II   & M0      & 1.5     & 1       & 5.69    & $-8.46$     & single                      & a \\
DR Tau     & II   & K5      & 1.2     & 3       & 5.26    & $-6.25$     & single                      & b \\
FT Tau     & II   & \nodata & \nodata & \nodata & \nodata & \nodata     & single                      & d \\
GM Aur     & II   & K3      & 1.2     & 2       & 5.95    & $-8.02$     & single                      & d \\
GO Tau     & II   & M0      & 2.2     & 2       & 6.25    & $-7.93$     & single                      & d \\
RY Tau     & II   & G1      & 2.2     & 4       & 6.88    & $-7.11$     & single                      & b \\
\hline
AS 205 (A) & II   & K5      & 2.9     & 5       & 5.00    & $-6.14$     & triple (1\farcs3, sb)       & b, e \\
AS 209     & II   & K5      & 0.9     & 6       & 5.48    & $-7.39$     & single                      & b \\
DoAr 25    & II   & K5      & 2.9     & 7       & 5.97    & $\le -9.24$ & single                      & f \\
DoAr 44    & II   & K3      & 2.1     & 8       & \nodata & \nodata     & single                      & b \\
Elias 24   & II   & K6      & 7.5     & 7       & 5.66    & $-6.67$     & single                      & f \\
GSS 39     & II   & M0      & 14      & 9       & 5.10    & $-7.20$     & single                      & f \\
L1709 B    & I/II & \nodata & \nodata & \nodata & \nodata & \nodata     & single (?)                  & \nodata \\
SR 21 (A)  & II   & G3      & 9.0     & 5       & 6.00    & $\le -8.84$ & binary (6\farcs7)           & g \\
SR 24 (S)  & II   & K1      & 5.5     & 7       & 6.64    & $-7.15$     & triple (5\farcs2, 0\farcs2) & g, a \\
WaOph 6    & II   & K6      & 3.5     & 10      & 5.78    & $-6.64$     & single (?)                  & \nodata \\
WSB 60     & II   & M4      & 2.0     & 7       & 6.34    & $-8.43$     & single                      & h \\
WL 20 (S)  & I    & \nodata & \nodata & \nodata & \nodata & \nodata     & triple (2\farcs2, 3\farcs6) & f 
\enddata
\tablecomments{Col.~(1): Object name.  Col.~(2): SED classification.  Col.~(3): 
Spectral type.  Col.~(4): Visual extinction.  Col.~(5): SpT and $A_V$ 
references are as follows: 1 - \citet{white04}; 2 - \citet{kh95}; 3 - 
\citet{mora01}; 4 - \citet{calvet04}; 5 - \citet{prato03}; 6 - 
\citet{herbig88}; 7 - \citet{wilking05}; 8 - \citet{bouvier92}; 9 - 
\citet{greene95}; 10 - \citet{eisner05}.  Col.~(6): The logarithm of the 
stellar age, from the literature.  Ages for AA Tau, CI Tau, DH Tau, DM Tau, DN 
Tau, GM Aur, and GO Tau are from \citet{hartmann98}.  Ages for DL Tau and DR 
Tau are from \citet{hartigan95}.  Other objects have ages from the same sources 
as the SpT references.  Col.~(7): The logarithm of the mass accretion rate, 
from the literature.  Accretion rates for AA Tau, CI Tau, DH Tau, DN Tau, GM 
Aur, and GO Tau are from \citet{hartmann98}; those for DoAr 25, Elias 
24, GSS 39, SR 21 (A), SR 24 (S), and WSB 60 are from \citet{natta06}.  The AS 
209 and DL Tau values are from \citet{valenti93} and the DR Tau value is from 
\citet{hartigan95}: these have been corrected as described by 
\citet[][their Table 4]{gullbring97}.  Others have \emph{\.{M}} from the same 
sources as the SpT references.  Col.~(8): Multiplicity and projected 
separations (sb refers to a spectroscopic binary system).  Col.~(9): 
Multiplicity references are as follows: a - \citet{simon95}; b - 
\citet{ghez93}; c - \citet{itoh05}; d - \citet{leinert93}; e - 
\citet{eisner05}; f - \citet{ratzka05}; g - \citet{reipurth93}; h - 
\citet{barsony03}.}
\end{deluxetable}

\clearpage

\begin{deluxetable}{lcccccc|rccc}
\tablecolumns{11}
\tabletypesize{\scriptsize}
\tablewidth{0pc}
\tablecaption{Disk Structure Parameters\label{structure_table}}
\tablehead{
\colhead{Object} & \colhead{$T_1$} & \colhead{$q$} & \colhead{$\Sigma_5$} & \colhead{$p$} & \colhead{$R_d$} & \colhead{$M_d$} & \colhead{$r_0$} & \colhead{$i$} & \colhead{PA} & \colhead{ref} \\
\colhead{} & \colhead{[K]} & \colhead{} & \colhead{[g cm$^{-2}$]} & \colhead{} & \colhead{[AU]} & \colhead{[M$_{\odot}$]} & \colhead{[AU]} & \colhead{[\degr]} & \colhead{[\degr]} & \colhead{} \\
\colhead{(1)} & \colhead{(2)} & \colhead{(3)} & \colhead{(4)} & \colhead{(5)} & \colhead{(6)} & \colhead{(7)} & \colhead{(8)} & \colhead{(9)} & \colhead{(10)} & \colhead{(11)}}
\startdata
04158+2805 & $189\,\,^{+15}_{-12}$ & $0.44\,\,^{+0.04}_{-0.02}$ & $0.3\,\,^{+0.4}_{-0.1}$ & $0.1\,\,^{+0.2}_{-0.1}$ & $700\pm100$             & $0.03\,\,^{+0.02}_{-0.01}$    & 0.10 & 68 & 88  & f, 1, 2 \\[0.2cm]
AA Tau     & $195\,\,^{+7}_{-15}$  & $0.59\,\,^{+0.07}_{-0.05}$ & $16\,\,^{+33}_{-6}$     & $0.9\,\,^{+0.7}_{-0.2}$ & $400\,\,^{+600}_{-75}$  & $0.03\,\,^{+0.07}_{-0.02}$    & 0.08 & 75 & 90  & 3, 4, 4 \\[0.2cm]
CI Tau     & $178\pm3$             & $0.57\pm0.02$              & $6\,\,^{+5}_{-3}$       & $0.3\pm0.3$             & $225\pm50$              & $0.04\pm0.01$                 & 0.10 & 46 & 131 & f, 2, 2 \\[0.2cm]
DH Tau (A) & $136\,\,^{+8}_{-5}$   & $0.63\,\,^{+0.04}_{-0.05}$ & $40\,\,^{+63}_{-39}$    & $1.0\pm1.0$             & $25\,\,^{+975}_{-25}$   & $0.003\,\,^{+0.282}_{-0.002}$  & 0.10 & 58 & 0   & f, 5, f \\[0.2cm]
DL Tau     & $172\pm2$             & $0.64\pm0.01$              & $40\,\,^{+49}_{-27}$    & $0.5\,\,^{+0.4}_{-0.5}$ & $175\,\,^{+50}_{-25}$   & $0.10\,\,^{+0.02}_{-0.01}$    & 0.10 & 35 & 44  & f, 6, 2 \\[0.2cm]
DM Tau     & 105\tablenotemark{a}  & 0.40\tablenotemark{a}      & $138\,\,^{+173}_{-134}$ & $1.7\,\,^{+0.3}_{-1.7}$ & $150\,\,^{+250}_{-100}$ & $0.014\,\,^{+0.003}_{-0.002}$ & 3.00 & 32 & 153 & 7, 6, 6 \\[0.2cm]
DN Tau     & $123\pm3$             & $0.64\pm0.02$              & $26\,\,^{+746}_{-14}$   & $0.2\,\,^{+1.8}_{-0.2}$ & $100\,\,^{+300}_{-25}$  & $0.06\pm0.02$                 & 0.07 & 28 & 40  & 8, 8, 2 \\[0.2cm]
DR Tau     & $315\pm5$             & $0.61\pm0.02$              & $11\,\,^{+142}_{-6}$    & $0.5\,\,^{+1.2}_{-0.5}$ & $100\,\,^{+175}_{-25}$  & $0.01\,\,^{+0.014}_{-0.002}$  & 0.15 & 72 & 170 & 8, 8, 2 \\[0.2cm]
FT Tau     & $147\,\,^{+8}_{-7}$   & $0.69\pm0.04$              & $299\,\,^{+1090}_{-259}$ & $1.0\pm1.0$             & $50\,\,^{+950}_{-25}$  & $0.05\pm0.03$                 & 0.10 & 60 & 82  & f, f, 9 \\[0.2cm]
GM Aur     & 130\tablenotemark{a}  & 0.43\tablenotemark{a}      & $51\,\,^{+124}_{-34}$   & $0.7\,\,^{+0.5}_{-0.4}$ & $150\pm25$              & $0.057\,\,^{+0.006}_{-0.002}$ & 24.0 & 55 & 58  & 7, 6, 6 \\[0.2cm]
GO Tau     & $130\,\,^{+3}_{-4}$   & $0.64\,\,^{+0.03}_{-0.02}$ & $9\,\,^{+44}_{-4}$      & $0.2\,\,^{+0.7}_{-0.2}$ & $350\,\,^{+650}_{-175}$ & $0.18\,\,^{+0.21}_{-0.11}$    & 0.10 & 66 & 0   & f, 2, 2 \\[0.2cm]
RY Tau     & $985\,\,^{+70}_{-68}$ & $0.70\pm0.04$              & $12\pm9$                & $0.5\,\,^{+0.2}_{-0.5}$ & $150\pm25$              & $0.02\pm0.01$                 & 0.54 & 86 & 32  & 8, 8, 2 \\[0.2cm]
\hline
\\[-0.13cm]
AS 205 (A) & $383\pm7$             & $0.65\pm0.03$              & $10\,\,^{+99}_{-3}$     & $0.1\,\,^{+1.0}_{-0.1}$ & $200\,\,^{+100}_{-25}$  & $0.11\,\,^{+0.02}_{-0.03}$    & 0.07 & 47 & 55  & 10, 2, 2 \\[0.2cm]
AS 209     & $247\,\,^{+7}_{-10}$  & $0.62\,\,^{+0.07}_{-0.03}$ & $12\,\,^{+4}_{-2}$      & $0.2\pm0.2$             & $200\pm25$              & $0.09\,\,^{+0.14}_{-0.02}$    & 0.10 & 27 & 35  & f, 2, 2 \\[0.2cm]
DoAr 25    & 125\tablenotemark{a}  & 0.40\tablenotemark{a}      & $4\,\,^{+5}_{-1}$       & $0.1\,\,^{+0.3}_{-0.1}$ & $200\pm25$              & $0.045\pm0.002$               & 0.10 & 62 & 112 & f, 2, 2 \\[0.2cm]
DoAr 44    & $229\pm5$             & $0.49\,\,^{+0.05}_{-0.04}$ & $60\,\,^{+88}_{-59}$    & $1.0\pm1.0$             & $25\,\,^{+975}_{-25}$   & $0.004\,\,^{+0.040}_{-0.001}$ & 0.10 & 60 & 0   & f, f, f \\[0.2cm]
Elias 24   & $278\pm7$             & $0.64\,\,^{+0.06}_{-0.04}$ & $21\,\,^{+2}_{-3}$      & $0.2\,\,^{+0.1}_{-0.2}$ & $175\pm25$              & $0.13\,\,^{+0.09}_{-0.03}$    & 0.10 & 45 & 27  & f, 2, 2 \\[0.2cm]
GSS 39     & $194\,\,^{+9}_{-12}$  & $0.55\,\,^{+0.07}_{-0.05}$ & $10\,\,^{+23}_{-6}$     & $0.2\,\,^{+0.5}_{-0.2}$ & $275\,\,^{+75}_{-25}$   & $0.13\,\,^{+0.25}_{-0.06}$    & 0.10 & 49 & 114 & f, 2, 2 \\[0.2cm]
L1709 B    & $234\pm2$             & $0.47\pm0.01$              & $71\,\,^{+249}_{-64}$   & $1.0\,\,^{+0.6}_{-1.0}$ & $225\,\,^{+125}_{-75}$  & $0.06\pm0.01$                 & 0.10 & 55 & 27  & f, 2, 2 \\[0.2cm]
SR 21 (A)  & $283\,\,^{+7}_{-11}$  & $0.49\,\,^{+0.14}_{-0.09}$ & $33\,\,^{+40}_{-13}$    & $1.4\pm0.5$             & $600\,\,^{+400}_{-250}$ & $0.02\,\,^{+0.10}_{-0.01}$    & 0.10 & 60 & 12  & f, f, 2 \\[0.2cm]
SR 24 (S)  & $236\,\,^{+7}_{-8}$   & $0.63\,\,^{+0.05}_{-0.04}$ & $46\,\,^{+35}_{-16}$    & $0.9\,\,^{+0.3}_{-0.2}$ & $500\,\,^{+500}_{-175}$ & $0.12\,\,^{+0.27}_{-0.05}$    & 0.10 & 57 & 25  & f, 2, 2 \\[0.2cm]
WaOph 6    & $173\pm5$             & $0.65\pm0.03$              & $69\,\,^{+55}_{-27}$    & $0.7\pm0.3$             & $275\,\,^{+225}_{-50}$  & $0.17\,\,^{+0.26}_{-0.06}$    & 0.02 & 41 & 12  & 10, 2, 2 \\[0.2cm]
WSB 60     & $118\,\,^{+4}_{-7}$   & $0.56\,\,^{+0.08}_{-0.03}$ & $59\,\,^{+70}_{-49}$    & $0.9\,\,^{+0.4}_{-0.8}$ & $350\,\,^{+650}_{-175}$ & $0.10\,\,^{+0.40}_{-0.04}$    & 0.10 & 42 & 3   & f, 2, 2 
\enddata
\tablecomments{Col~(1): Object name.  Col~(2): Temperature at 1\,AU.  Col.~(3): 
Power-law index of radial temperature distribution.  Col.~(4): Surface density 
at 5\,AU (gas + dust).  Col.~(5): Power-law index of radial surface density 
distribution.  Col.~(6): Outer radius.  Col.~(7): Total disk mass.  Col.~(8): 
Inner radius (fixed in minimization).  Col.~(9): Inclination (fixed in 
minimization).  Col.~(10): Position angle (fixed in minimization).  Col.~(11): 
References for \{$r_0$, $i$, PA\} as follows: f - fixed, assumed value; 1 - 
F.~M\'{e}nard 2005, private communication; 2 - this paper (see Table 
\ref{cont_table}); 3 - \citet{bouvier99}; 4 - \citet{osullivan05}; 5 - 
\citet{bouvier95}; 6 - \citet{simon00}; 7 - \citet{calvet05}; 8 - 
\citet{muzerolle03}; 9 - \citet{dutrey96}; 10 - \citet{eisner05}.}
\tablenotetext{a}{$T_r$ parameters are fixed in the minimizations, and were 
derived from simple power-law approximations to the midplane temperatures 
calculated by \citet{dalessio05}, as described in the text.  Infrared SEDs are 
not used in these fits.}
\end{deluxetable}

\begin{deluxetable}{lccc}
\tablecolumns{4}
\tabletypesize{\scriptsize}
\tablewidth{0pc}
\tablecaption{Opacity Constraints\label{opacity_table}}
\tablehead{
\colhead{Object} & \colhead{$n$} & \colhead{$\Delta$} & \colhead{$\beta$} \\
\colhead{(1)} & \colhead{(2)} & \colhead{(3)} & \colhead{(4)}}
\startdata
04158+2805 & $2.7\pm0.4$ & 0.0     & $0.7\pm0.4$ \\
AA Tau     & $2.8\pm0.3$ & 0.3     & $1.0\pm0.3$ \\
CI Tau     & $2.9\pm0.2$ & 0.0     & $0.9\pm0.2$ \\
DH Tau     & $1.8\pm0.4$ & \nodata & \nodata     \\
DL Tau     & $2.8\pm0.1$ & 0.3     & $1.0\pm0.1$ \\
DM Tau     & $2.9\pm0.2$ & 0.7     & $1.5\pm0.2$ \\
DN Tau     & $2.3\pm0.2$ & 1.4     & $0.7\pm0.2$ \\
DR Tau     & $2.1\pm0.3$ & 0.6     & $0.2\pm0.3$ \\
FT Tau     & $2.6\pm0.2$ & \nodata & $\ge 0.6$   \\
GM Aur     & $3.1\pm0.1$ & 0.1     & $1.2\pm0.1$ \\
GO Tau     & $1.8\pm0.7$ & 0.5     & \nodata     \\
RY Tau     & $2.7\pm0.1$ & \nodata & $\ge 0.7$   \\
\hline
AS 205 (A) & $2.3\pm0.4$ & 0.3     & $0.4\pm0.4$ \\
AS 209     & $1.5\pm0.7$ & 0.3     & \nodata     \\
DoAr 25    & $1.7\pm0.3$ & 0.0     & \nodata     \\ 
DoAr 44    & $1.4\pm0.6$ & \nodata & \nodata     \\
Elias 24   & $2.1\pm0.3$ & 0.1     & $0.1\pm0.3$ \\
GSS 39     & $2.5\pm0.5$ & 0.1     & $0.6\pm0.5$ \\
L1709 B    & $2.6\pm0.4$ & 0.4     & $0.8\pm0.4$ \\
SR 21 (A)  & $2.4\pm0.7$ & 0.3     & $0.5\pm0.7$ \\
SR 24 (S)  & $2.1\pm0.6$ & 0.3     & $0.1\pm0.6$ \\
WaOph 6    & $2.6\pm0.7$ & 0.5     & $0.9\pm0.7$ \\
WSB 60     & $1.6\pm0.2$ & 0.4     & \nodata     \\
WL 20 (S)  & $2.5\pm0.2$ & \nodata & $\ge 0.5$   
\enddata
\tablecomments{Col.~(1): Object name.  Col.~(2): Best-fit power-law index $n$ 
of the form $F_{\nu} \propto \nu^n$, for wavelengths longer than either 850 or 
1000\,$\mu$m (depending on the available data).  Col.~(3): Maximum allowed 
value of $\Delta$, the ratio of optically thick to thin emission at the 
shortest wavelength used to determine $n$, as described in the text.  Col.~(4): 
Corresponding value of the power-law index of the opacity spectrum, $\beta$ 
(see text).}
\end{deluxetable}

\clearpage

\clearpage

\clearpage

\clearpage

\clearpage

\clearpage

\clearpage
                                                                                
\clearpage

\clearpage

\clearpage

\clearpage

\clearpage

\clearpage

\clearpage

\clearpage

\clearpage

\clearpage

\clearpage

\begin{deluxetable}{lccccccccccccccc}
\rotate
\tablecolumns{16}
\tabletypesize{\scriptsize}
\tablewidth{0pc}
\tablecaption{Literature Sources for SEDs\label{seds_table}}
\tablehead{
\colhead{Object} & \multicolumn{15}{c}{References for $F_{\nu}$ at Various Wavelengths (in $\mu$m)} \\ \colhead{} & \colhead{8} & \colhead{10} & \colhead{12} & \colhead{25} & \colhead{60} & \colhead{350} & \colhead{450} & \colhead{600} & \colhead{800} & \colhead{850} & \colhead{1000} & \colhead{1300} & \colhead{2000} & \colhead{2700} & \colhead{7000}}
\startdata
04158+2805 & \nodata & \nodata & 1       & 1       & 1       & 2       & 2       & \nodata & \nodata & 2       & \nodata & 3       & \nodata & \nodata & \nodata \\
AA Tau     & \nodata & 4       & 5       & 5       & 5       & 6       & 6       & \nodata & 7       & 6       & 7       & 8       & 9       & 10      & \nodata \\
CI Tau     & \nodata & 11      & 5       & 5       & 5       & 6       & 6       & 7       & 7       & 6       & 7       & 8       & \nodata & 10      & 12      \\
DH Tau (A) & \nodata & \nodata & 5       & 5       & 5       & 6       & 6       & \nodata & 13      & 6       & \nodata & \nodata & \nodata & 10      & \nodata \\
DL Tau     & 14      & \nodata & 5       & 5       & 5       & 15      & 15      & 15, 7   & 7, 15   & 15      & 7, 15   & 8       & 9       & 10      & 12      \\
DM Tau     & \nodata & \nodata & \nodata & \nodata & \nodata & 6       & \nodata & 7       & 7       & 6       & 7       & 8       & 9       & 10      & 12      \\
DN Tau     & \nodata & 16      & 5       & 5       & 5       & 6       & \nodata & \nodata & 7       & 6       & 7       & 8       & 9       & 10      & \nodata \\
DR Tau     & 14      & 11      & 5       & 5       & 5       & 15      & 6, 15   & 15      & 7, 15   & 15, 6   & 7, 15   & 8       & 9       & \nodata & \nodata \\
FT Tau     & \nodata & \nodata & 5       & 5       & 5       & 6       & 6       & 7       & 7       & 6       & 7       & 8       & \nodata & 10      & 12      \\
GM Aur     & \nodata & \nodata & \nodata & \nodata & \nodata & 6       & 17      & 7       & 7, 17   & \nodata & 7       & 8       & 9       & 18      & 12      \\
GO Tau     & 14      & 16      & 5       & 5       & 5       & 6       & \nodata & \nodata & \nodata & 6       & \nodata & 8       & \nodata & \nodata & \nodata \\
RY Tau     & \nodata & 11      & 5       & 5       & 5       & 15      & 15      & 15, 7   & 7, 15   & 15      & 7, 15   & 8       & 15, 9   & \nodata & 12      \\
\hline
AS 205 (A) & 19      & 19      & 5, 19   & 5       & 5       & 20      & 20      & \nodata & 21      & 20      & 21      & 22      & \nodata & \nodata & \nodata \\
AS 209     & \nodata & 19      & 5       & 5       & 5       & 20      & 20      & \nodata & \nodata & 20      & \nodata & 22      & \nodata & \nodata & \nodata \\
DoAr 25    & \nodata & \nodata & \nodata & \nodata & \nodata & \nodata & 23      & \nodata & 23      & 20      & \nodata & 22      & \nodata & \nodata & \nodata \\
DoAr 44    & \nodata & \nodata & 24      & 24      & 25      & \nodata & \nodata & \nodata & \nodata & 20      & \nodata & 26      & \nodata & \nodata & \nodata \\
Elias 24   & \nodata & 27      & 24, 28  & 24      & \nodata & 20      & 20      & \nodata & \nodata & 20      & \nodata & 22      & \nodata & \nodata & \nodata \\
GSS 39     & \nodata & 27      & 29      & \nodata & \nodata & \nodata & 20      & \nodata & \nodata & 20      & \nodata & 22      & \nodata & \nodata & \nodata \\
L1709 B    & \nodata & \nodata & 24      & 24      & 24      & 20      & 20, 23  & \nodata & 23      & 20      & \nodata & 22      & \nodata & \nodata & \nodata \\
SR 21 (A)  & \nodata & 27      & 24, 28  & \nodata &         & 20      & 20      & \nodata & \nodata & 20      & \nodata & 22      & \nodata & \nodata & \nodata \\
SR 24 (S)  & \nodata & 30      & \nodata & 30      & \nodata & 20      & 20      & \nodata & \nodata & 20      & \nodata & 26      & \nodata & \nodata & \nodata \\
WaOph 6    & 31, 32  & 31      & \nodata & 31      & 31      & \nodata & 20      & \nodata & \nodata & 20      & \nodata & 22      & \nodata & \nodata & \nodata \\
WSB 60     & \nodata & 33      & 33, 28  & 33      & \nodata & 20      & 20      & \nodata & \nodata & 20      & \nodata & 22      & \nodata & \nodata & \nodata 
\enddata
\tablecomments{References are as follows: 1 - \citet{kenyon90}; 2 - M. C. Liu, 
private communication; 3 - \citet{motte01}; 4 - \citet{metchev04}; 5 - 
\citet{weaver92}; 6 - \citet{andrews05}; 7 - \citet{beckwith91}; 8 - 
\citet{bscg90}; 9 - \citet{kitamura02}; 10 - \citet{dutrey96}; 11 - 
\citet{kh95}; 12 - \citet{rodmann06}; 13 - \citet{jewitt94}; 14 - 
\citet{hartmann05}; 15 - \citet{mannings94}; 16 - \citet{simon95b}; 17 - 
\citet{weintraub89}; 18 - \citet{looney00}; 19 - \citet{liu96}; 20 - Andrews \& 
Williams, in preparation; 21 - \citet{jensen96}; 22 - \citet{andre94}; 23 - 
\citet{dent98}; 24 - \emph{IRAS} Point Source Catalog; 25 - \citet{clarke91}; 
26 - \citet{nurnberger98}; 27 - \citet{barsony05}; 28 - \citet{bontemps01}; 29 
- \citet{lada84}; 30 - \citet{mccabe06}; 31 - \citet{padgett06}; 32 - 
\citet{gras-velazquez05}; 33 - \citet{wilking89}.}
\end{deluxetable}

\end{document}